\documentclass[12pt]{iopart}
\newcommand{\be}{\begin{equation}}
\newcommand{\ee}{\end{equation}}
\newcommand{\bea}{\begin{eqnarray}}
\newcommand{\eea}{\end{eqnarray}}
\def\alt{\stackrel{<}{\sim}}
\def\agt{\stackrel{>}{\sim}}
\def\eslt{E_T^{\rm miss}}

\def\to{\rightarrow}

\def\bi{\begin{itemize}}
\def\ei{\end{itemize}}
\def\te{\tilde e}
\def\tG{\tilde G}
\def\ta{\tilde a}

\def\tst{\tilde t}
\def\ttau{\tilde \tau}

\def\tg{\tilde g}

\def\tq{\tilde q}

\def\tw{\widetilde W}
\def\tz{\widetilde Z}

\newcommand\prd[3]{{\it Phys.\ Rev.\ }{\bf D #1} (#2) #3}
\newcommand\prep[3]{{\it Phys.\ Rept.\ }{\bf #1} (#2) #3}
\newcommand\prl[3]{{\it Phys.\ Rev.\ Lett.\ }{\bf #1} (#2) #3}
\newcommand\plb[3]{{\it Phys.\ Lett.\ }{\bf B #1} (#2) #3}
\newcommand\jhep[3]{{\it J. High Energy Phys.\ }{\bf #1} (#2) #3}
\newcommand\app[3]{{\it Astropart.\ Phys.\ }{\bf #1} (#2) #3}
\newcommand\apj[3]{{\it Astrophys.\ J. }{\bf #1} (#2) #3}
\newcommand\ijmpd[3]{{\it Int.\ J.\ Mod.\ Phys.\ }{\bf D #1} (#2) #3}
\newcommand\npb[3]{{\it Nucl.\ Phys.\ }{\bf B #1} (#2) #3}
\newcommand\epjc[3]{{\it Eur.\ Phys.\ J. }{\bf C #1} (#2) #3}
\newcommand\ptp[3]{{\it Prog.\ Theor.\ Phys.\ }{\bf #1} (#2) #3}

\newcommand\arnps[3]{{\it Ann.\ Rev.\ Nucl.\ Part.\ Sci.}{\bf  #1} (#2) #3}

\newcommand\jphg[3]{{\it J.\ Phys.}{\bf G #1} (#2) #3}

\newcommand{\hepph}[1]{hep-ph/#1}
\newcommand{\astroph}[1]{astro-ph/#1}
\usepackage{graphicx}
\begin{document}

\title{Collider, direct and indirect detection
 of supersymmetric dark matter}

\author{Howard Baer}
\address{Dep't of Physics and Astronomy, University of Oklahoma, 
Norman, OK, 73019, USA}
\ead{baer@nhn.ou.edu}

\author{Eun-Kyung Park} 
\address{Physikalisches Institut, Universit$\ddot{a}$t Bonn, Nussallee 12, D53115 Bonn, Germany}
\ead{epark@th.physik.uni-bonn.de}

\author{Xerxes Tata} 
\address{Dept. of Physics and Astronomy, University of Hawaii, Honolulu, HI 96822, USA}
\ead{tata@phys.hawaii.edu}

\begin{abstract}
We present an overview of supersymmetry searches, both at collider
experiments and via searches for dark matter (DM).  We focus on three DM
possibilities in the SUSY context: the thermally produced neutralino, a
mixture of axion and axino, and the gravitino, and compare and contrast
signals that may be expected at colliders, in direct detection
(DD) experiments searching of DM relics left over from the Big Bang, and
indirect detection (ID) experiments designed to detect the products of
DM annihilations within the solar interior or galactic halo. Detection
of DM particles using multiple strategies  provides
complementary information that may shed light on the new physics
associated with the dark matter sector.  In contrast to the mSUGRA model
where the measured cold DM relic density restricts us to special
regions mostly on the edge of the $m_0-m_{1/2}$ plane, the entire
parameter plane becomes allowed if the universality assumption is
relaxed in models with just one additional parameter. Then, thermally produced
neutralinos with a well-tempered mix of wino, bino and higgsino
components, or with a mass adjusted so that their annihilation in the early
universe is Higgs-resonance-enhanced, can be the DM.  Well-tempered
neutralinos typically yield heightened rates for DD and ID experiments
compared to generic predictions from minimal supergravity. If instead DM
consists of axinos (possibly together with axions) or gravitinos, then
there exists the possibility of detection of quasi-stable
next-to-lightest SUSY particles at colliding beam experiments, with
especially striking consequences if the NLSP is charged, but no DD
or ID detection. The exception for mixed axion/axino DM is that DD of
axions may be possible.

\end{abstract}

\maketitle

\section{Introduction: dark matter in SUSY models}
\label{sec:intro}

In the 1930's, the astronomer Fritz Zwicky noticed something was amiss
in the universe\cite{zwicky}: observations of galactic clusters seemed
in contradiction with the amount of luminous matter present in
them. Specifically, 
they seemed to be lacking enough gravitational pull in order to maintain
themselves as bound clusters.  To account for the discrepancy without
modifying the laws of gravity, Zwicky hypothesized that most of the mass
of the galaxy was contained in
non-luminous, or dark, matter (DM). Few paid attention to Zwicky's
hypothesis until the 1970s, when Ford and Rubin, measuring the rotation
curves of galaxies\cite{rubin}, found that stellar velocities did not
drop-off with radial distance in accord with Newton's laws, but instead
stayed high out to the largest distances accessible to observation. An
explanation could be found by resuscitating Zwicky's DM
conjecture.

In recent times, cosmology has entered a much more quantitative period,
highlighted by: detailed measurements of anisotropies in the cosmic
microwave background radiation\cite{wmap}, measurements of galactic
lensing\cite{lense} and comparisons of large scale structure to $n$-body
simulations of the development of structure in the universe\cite{nbody}.
All these measurements, when combined into a standard cosmological
model, point decisively towards a universe constituted of 4\% baryonic
matter, along with $\sim 25\%$ cold dark matter, and about 70\% dark
energy (DE)\cite{comp}.  A tiny fraction remains associated with
electrons, neutrinos and photons. The accelerating Universe
\cite{accel}, and the concomitant DE, came as a surprise in the late
1990s. A cosmological constant was not unanticipated in theoretical
cosmology, and an upper bound nearly equal to its measured value
was obtained a decade earlier\cite{weinberg_cc}. 
Although the origin of DE remains an
outstanding puzzle, much mystery remains around the DM as
well\cite{cdm_reviews}. While the amount of DM in the universe is
becoming ever-more precisely known, the identity of the particle (or
particles) is completely unknown. What is known is that the bulk of the
dark matter must be {\it cold}, {\it i.e.} non-relativistic particles
with velocities so low they can clump, or become gravitational bound on
large scales, thus providing the seeds for structure formation.
This rules out active neutrinos as DM.
Unraveling the nature of the cold dark matter (CDM) in the universe is
one of the most exciting directions in scientific research today.
Happily, a bevy of experiments currently operating, being deployed, or
in the planning stage, promises rapid progress on uncovering the
properties of CDM during the next few years.

{\it None} of the particles of the Standard Model (SM) of particle
physics (which encapsulates the laws of physics as we know them) has the
right properties to make up the CDM, calling for a major revision in our
knowledge of the laws of physics. Indeed the SM is best viewed as an
{\it effective field theory}, a set of laws that gives a valid
description of nature {\it up to the weak interaction energy scale}
$\sim 0.1-1$ TeV; it almost certainly breaks down beyond this scale, as
evidenced by instabilities in the electroweak symmetry breaking sector
of the theory.

On the cosmology side, if one assumes the existence of a DM particle
that was in thermal equilibrium early in the universe's history, and has
not been produced after the Universe cooled below the DM particle mass,
one can unambiguously calculate its relic abundance by solving its
Boltzmann equation. The answer depends on the dark matter particle's
annihilation cross section and mass. Remarkably, a DM particle with a
weak scale mass and an annihilation cross section of weak interaction
size yields about the observed relic density, strongly suggesting a {\it
weakly interacting massive particle}, or WIMP, as the CDM candidate
(though other possibilities also exist\cite{fk}.).  This is often
pointed to as providing independent astrophysical evidence that new
physics ought to exist at the weak scale, and is sometimes termed {\it
the WIMP miracle}.  The goal of the CERN Large Hadron Collider
experiments -- which will begin gathering data starting in late 2009 --
is to make a thorough exploration for new matter states and interactions
in and around the weak scale.

The theoretical literature is replete with candidate CDM particles.
While some of these are postulated specifically to solve the CDM
problem, others emerge as solutions to long-standing problems in
particle physics. Examples of the latter include {\it axions}, which
emerge from the Peccei-Quinn (PQ) solution to the strong $CP$
problem\cite{pqww}, and WIMPs, that are frequently contained in particle
physics theories that attempt to stabilize the weak scale. In this
paper, we will focus upon dark matter candidates which emerge from
particle physics models with {\it weak scale supersymmetry} (SUSY)
\cite{wss}. In the Minimal Supersymmetric Standard Model (MSSM) with a
conserved $R$-parity,
the lightest SUSY particle (LSP) is absolutely stable.
In many SUSY models, the LSP is the massive, electrically neutral, and
weakly interacting lightest neutralino $\tz_1$, and thus an excellent
WIMP candidate. If one includes the gravity multiplet -- including
the graviton and spin-${3\over 2}$ gravitino states -- then the
gravitino $\tG$ is also a good CDM candidate. In this case, since $\tG$
only interacts gravitationally, it is usually termed a {\it
superWIMP}\cite{feng}.  Finally, in models where the PQ solution to the
strong $CP$ problem is invoked, spin-0 axions and their $R$-odd
spin-${1\over 2}$ partner {\it axinos} $\ta$ occur. In this case, both
the axion\cite{absik} and axino\cite{wilc} can account for the CDM. The
axino is sometimes called an extremely weakly interacting massive
particle, or {\it eWIMP} \cite{roszk}.\footnote{A fourth SUSY CDM
candidate, the right-handed sneutrino, is also possible: see
Ref. \cite{moroi_rhsn} for further details.} Weak scale SUSY models
{\it i}). solve the hierarchy problem, 
{\it ii}). naturally accommodate CDM, and
{\it iii}). automatically lead to the unification of the 
measured gauge couplings, a triple coincidence that seems hard to ignore.

If CDM is dominantly WIMPs, then it may be possible to produce and
study the DM particle(s) directly at colliding beam experiments such as
the CERN LHC. Direct production of DM particles is not likely to
be visible above SM backgrounds at LHC. However, 
production of new matter states {\it associated with the DM}, and which decay into DM particles, often lead to robust new physics signatures.
In such scenarios the LHC may then turn out to be a DM factory,
where the nature of DM particles and their properties might be studied
in a controlled environment.  In a collider detector, WIMPs would be
like neutrinos in that they would escape without depositing
any energy in the experimental apparatus, resulting in an {\it apparent
imbalance of energy and momentum} in collider events. While WIMPs would
manifest themselves only as {\it missing (transverse) energy} in
(hadron) collider experiments, it should nevertheless be possible to study the
{\it visible} particles produced in WIMP-related production and decay
processes to study the new physics associated with the WIMP sector.

Indeed, there exists a real possibility that the nature of WIMP DM and
its associated new particle sector will be clarified in the next decade
by a {\it variety} of experiments that are already operating, or are
soon-to-be deployed.  In this effort, experiments at the LHC will play a
crucial role.  There are -- in tandem with the LHC -- a variety of other
dark matter search experiments already in operation, or in a deployment
or planning phase.  {\it Direct detection} (DD) experiments seek to
directly measure relic DM particles left over from early stages of the
Big Bang.  These DD experiments range from terrestrial microwave
cavities that search for axions via their conversion to photons, to
crystalline or noble liquid targets located deep underground that allow
for a search for WIMP-nucleon collisions by detecting the nuclear
recoil.

DM may also be searched for in {\it indirect detection} (ID)
experiments. In ID experiments, one searches for WIMP-WIMP annihilation
into various SM particles including neutrinos, gamma rays and
anti-matter.  Clearly, this technique applies only if the DM is
self-conjugate, or if DM particles and anti-particles are roughly
equally abundant.  One ID search method involves the use of neutrino
telescopes mounted deep under water or in polar ice. The idea is that if
relic WIMPs are the DM in our galactic halo, the sun (or earth) will
sweep them up as it follows its galactic orbit. The WIMPs then become
gravitationally trapped in the solar core where they can accumulate,
essentially at rest, to densities much higher than in the Milky Way
halo.  These accumulated WIMPs can then annihilate one with another into
SM particles with energies $E\alt m_{\rm WIMP}$.  Most annihilation products
would be immediately absorbed by the solar material. However, neutrinos
produced as primaries or secondaries by WIMP annihilation, 
can easily escape the sun resulting in
an isotropic flux of {\it high energy} neutrinos from the solar core,
some of which would make it to earth.  For $m_{\rm WIMP} \ge$~few~GeV,
the resulting neutrino energies
are impossible to produce via conventional nuclear reactions in the sun.
The neutrinos will occasionally interact with nuclei in ocean water or
ice and convert to a high energy muon, which could then be detected via
Cerenkov radiation by photomultiplier tubes 
located within the medium.

Another possibility for ID is to search for the by-products of WIMP
annihilation in various regions of our galactic halo. Even though the
halo number density of WIMPs would be quite low, the volume of the
galaxy is enormous, and one can look for rare anti-matter production or
high energy gamma ray production from these WIMP halo annihilations. A
variety of land-based, high altitude and space-based anti-matter and
gamma ray detectors have been or are being deployed. The space-based
Pamela experiment is searching for positrons and anti-protons. The
land-based HESS telescope has recently been joined by the Fermi
Gamma-ray Space Telescope (FGST) in the search for high energy gamma
rays. While high energy anti-particles would provide a striking signal,
these lose energy upon deflection when traversing the complicated
galactic magnetic field, and so can only be detected over limited
distances.  Gamma rays, on the other hand, are undeflected by magnetic
fields, and so have an enormous range and, furthermore, point back
to their source.  Thus, the galactic center, where dark matter
is expected to accumulate at a high density, might be a good source of
GeV-scale gamma rays resulting from WIMP-WIMP annihilation to vector
boson ($V=W,Z$) pairs or to quark jets, followed by $(V\to) q\to\pi^0\to
\gamma\gamma$ after hadronization and decay.

If WIMPs and their associated particles are discovered at the LHC and/or
at DD or ID search experiments, it will be a revolutionary
discovery. But it will only be the beginning of the story as it will
usher in a new era of {\it dark matter astronomy}.  The next logical
step would be the construction of an $e^+e^-$ collider of sufficient
energy so that WIMP (and related particles) can be produced and studied
with high precision in a clean, well-controlled experimental
environment. The precise determination of particle physics quantities
associated with WIMP physics can allow us to {\it deduce} the 
expected WIMP relic
density within the standard Big Bang cosmology. If this
turns out to be in agreement with the measured relic density, we would
have direct evidence that DM consists of a single component. If the
predicted relic density is too small, it could make the case for
multiple components in the DM sector. If the predicted density is too
large, we would be forced to abandon the simplest picture and seek more
complicated (non-thermal) mechanisms to account for the measurement. In
this case, we would also be able to deduce that the detected WIMP
is itself unstable, and that 
the DM is perhaps some lighter decay product.  
The determination of the properties
of the DM sector will also serve as a tool for a detailed measurement of
astrophysical quantities such as the galactic and local WIMP density and
local velocity profiles, which could shed light on the formation of
galaxies and on the evolution of the universe.

\section{Neutralino dark matter in gravity-mediated SUSY breaking models}\label{sec:sugra}

Even with the assumption of $R$-parity conservation, the MSSM has a very
large number of parameters making phenomenological analyses
intractable. It is customary to make assumptions based on physical insight 
as to how SUSY breaking effects are communicated from the SUSY breaking sector 
to the SM superpartners. This has led to
the development of simple models, each with just a handful of
parameters, characterized by the mediation-mechanism for SUSY breaking,
and with distinct predictions for the masses and couplings of
sparticles. While these various models can all accommodate the observed
relic density, gravity-mediated SUSY breaking models lead to thermal
WIMP dark matter in the most natural way, and hence are the focus of our
attention.

Once a SUSY model is specified, then given a set of input parameters, it
is possible to compute all superpartner masses and couplings necessary
for phenomenology.  We can then use these to calculate scattering cross
sections and sparticle decay modes to evaluate SUSY signals (and
compare against corresponding SM backgrounds) in collider experiments.  We can also
check whether the model is allowed or excluded by experimental
constraints, either from direct SUSY searches, {\it e.g.} at LEP2 which
requires that $m_{\tw_1}>103.5 $ GeV, $m_{\te}\agt 100$~GeV, and
$m_h>114.4$ GeV (for a SM-like light SUSY Higgs boson $h$), or from
indirect searches through loop effects from SUSY particles in low energy
measurements such as $BF(b\to s\gamma)$ or $(g-2)_\mu$.  We can also
calculate the expected lightest neutralino relic density
$\Omega_{\tz_1}h^2$, assuming $\tz_1$ is the LSP, or for that matter any
other stable particle in the theory.  For the sparticle mass spectrum, we adopt the 
Isasugra subprogram of Isajet\cite{isajet}, while for the neutralino relic density
calculation, we adopt the IsaReD\cite{isared} subprogram; the latter
includes all relevant neutralino annihilation and co-annihilation reactions.

\subsection{The mSUGRA model}

The minimal supergravity model (mSUGRA)\cite{msugra}\footnote{This is often also
referred to as the constrained MSSM, or CMSSM, in the literature.} 
is a prototypical model for investigations of the phenomenological consequences of
weak scale supersymmetry.  The parameter space of the model is given by
\be
m_0,\ m_{1/2},\ A_0,\ \tan\beta ,\ sign(\mu ),
\ee
where $m_0$ is a common GUT scale soft SUSY breaking (SSB) scalar mass,
$m_{1/2}$ is a common GUT scale SSB gaugino mass, $A_0$ is a common GUT
scale trilinear SSB term, $\tan\beta$ is the ratio of Higgs field vevs,
and $\mu$ is the superpotential Higgs mass term, whose magnitude, but
not sign, is constrained by the electroweak symmetry breaking
minimization conditions.

\begin{figure}[tbh]
\begin{center}
\includegraphics[width=12cm]{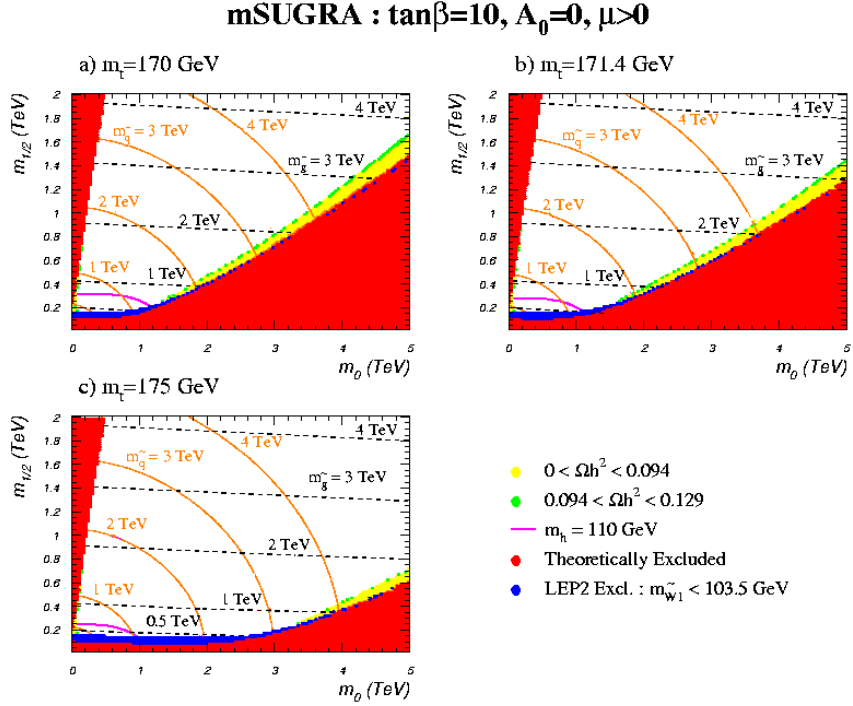}
\end{center}
\vspace{-3mm}
\caption{\small\it 
The $m_0\ vs.\ m_{1/2}$ plane in
mSUGRA for $A_0=0$, $\tan\beta =10$ with $\mu >0$ and {\it a})
$m_t=170$~GeV, {\it b}) $m_t=171.4$~GeV and {\it c}) $m_t=175$~GeV.  The
red-shaded regions are excluded because electroweak symmetry is not
correctly broken, or because the LSP is
charged. Blue regions are excluded by direct SUSY searches at LEP2.
Yellow and green shaded regions are WMAP-allowed, while white regions
are excluded owing to $\Omega_{\tz_1}h^2>0.129$.  Also shown are
gluino and first generation squark mass contours, as well as a magenta
contour below which $m_h\leq 110$~GeV.
}
\label{fig:msug} 
\end{figure}

To illustrate how various theoretical and experimental constraints constrain
the parameter space of the mSUGRA model, we show in Fig. \ref{fig:msug}
the $m_0\ vs.\ m_{1/2}$ plane, where we take
$A_0=0$, $\mu >0$ and $\tan\beta =10$ for three different values of
$m_t$.  The red-shaded regions are not allowed because either the
$\ttau_1$ becomes the lightest SUSY particle, in contradiction to
negative searches for long lived, charged relics (left edge), or EWSB is
not correctly obtained (lower-right region). The blue-shaded region is
excluded by LEP2 searches for chargino pair production
($m_{\tw_1}<103.5$ GeV).  Below the magenta contour near $m_{1/2}\sim
300$~GeV, $m_h<110$ GeV, which is roughly the LEP2 lower limit on $m_h$
in the model.  The thin green regions at the boundary of the unshaded white
region has $0.094<\Omega_{\tz_1}h^2<0.129$ where the neutralino
saturates the observed relic density.  In the adjoining yellow regions,
$\Omega_{\tz_1}h^2<0.094$, so these regions require multiple DM
components. The white regions all have $\Omega_{\tz_1}h^2>0.129$ and so
give too much thermal DM: they are excluded in the standard Big Bang
cosmology unless the neutralino decays either via small $R$-parity
violating couplings, or the model is extended to include yet lighter
sparticles.\footnote{For non-standard cosmology, 
then all bets are off \cite{gg}.} 
For the reader's convenience, we also show contours of
constant gluino and first generation squark mass, which  are useful for
understanding the SUSY reach of the LHC.

The DM-allowed regions are classified as follows:
\bi
\item At very low $m_0$ and low $m_{1/2}$ values is the so-called {\it
bulk} annihilation region\cite{bulk}.  Here, sleptons are quite light,
so $\tz_1\tz_1\to \ell\bar{\ell} $ via $t$-channel slepton exchange is
the dominant neutralino annihilation process in the early universe.

\item At low $m_0$ and moderate $m_{1/2}$, there is a thin strip of
allowed region adjacent to the stau-LSP region where the neutralino and
the lighter stau were in thermal equilibrium in the early universe. Here,
neutralino co-annihilation with the light stau serves to bring the
neutralino relic density down to its observed value\cite{stau}.

\item At large $m_0$, adjacent to the EWSB excluded region on the right,
is the hyperbolic branch/focus point (HB/FP) region, where the
superpotential $\mu$ parameter becomes small and the higgsino-content of
$\tz_1$ increases significantly. Then $\tz_1$ becomes mixed
higgsino-bino dark matter  (MHDM) and can annihilate efficiently via the
gauge coupling to its higgsino component. If $m_{\tz_1}>M_W$ and $M_Z$,
then $\tz_1\tz_1\to WW,\ ZZ,\ Zh$ is enhanced, and one finds the correct
measured relic density\cite{hb_fp}. Deep in the HB/FP region,
co-annihilation with the (higgsino-like) $\tw_1$ and $\tz_2$ can be
important. 

\ei

\begin{figure}[tbh]
\begin{center}
\includegraphics[width=10cm]{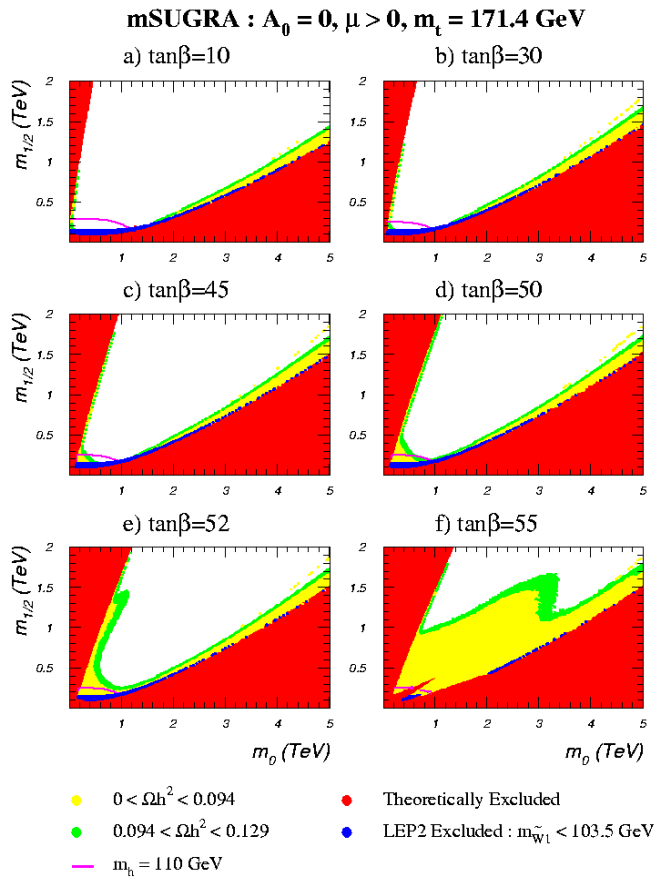}
\end{center}
\vspace{-3mm}
\caption{\small\it 
The $m_0\ vs.\ m_{1/2}$ plane in mSUGRA for $A_0=0$
and various values of $\tan\beta$, with $\mu >0$ and $m_t=171.4$~GeV.
The red-shaded regions are excluded because electroweak symmetry is not
correctly broken, or because the 
LSP is charged. Blue regions are excluded by direct SUSY searches at LEP2.
Yellow and green shaded regions are WMAP-allowed, while white 
regions are excluded owing to $\Omega_{\tz_1}h^2>0.129$. Below the
magenta contour in each frame, $m_h< 110$~GeV. 
}
\label{fig:sugpmu} 
\end{figure}

If the parameter $\tan\beta$ is increased much beyond 10, then
bottom and tau Yukawa couplings become large, and the value of
$m_A$ steadily drops. The situation is depicted in Fig. \ref{fig:sugpmu},
where we show the mSUGRA $m_0\ vs.\ m_{1/2}$ plane for increasing values of
$\tan\beta$.
\bi
\item For $\tan\beta\sim 45-55$, the value of $m_A$ is small enough
so that $\tz_1\tz_1$ can annihilate into $b\bar{b}$ pairs through
the $s$-channel $A$ (and also $H$) resonance.  This
region has been dubbed the $A$-funnel\cite{Afunnel}. 
It can be quite broad at large
$\tan\beta$ because the width $\Gamma_A$ becomes very wide due to 
the large $b$- and $\tau$- Yukawa  couplings.
\item It is also possible at low $m_{1/2}$ values that a
light Higgs $h$ resonance annihilation region can occur just above the
LEP2 excluded region\cite{hfunnel}.  
\item Finally, if $A_0$ is large and negative, then the $\tst_1$ can
become light. If $m_{\tst_1}\sim m_{\tz_1}$, then stop-neutralino
co-annihilation\cite{stop_co} can occur.  \ei 
Bino-wino coannihilation,
which is possible in extended models discussed below, is not possible in
this model on account of the assumed unification of gaugino mass
parameters.

\subsection{Direct and indirect detection of neutralino DM}

\begin{figure}[tbh]
\begin{center}
\includegraphics[width=7cm]{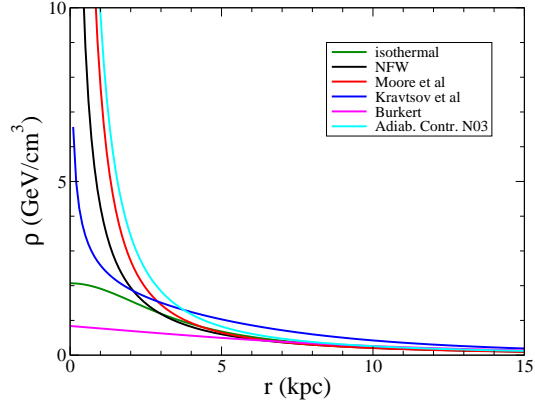}
\end{center}
\vspace{-3mm}
\caption{\small\it Several galactic dark matter
  halo density profiles. Notice that while these differ greatly close to
  the core of our Galaxy, they agree at the location
  of the sun, about 8~kpc from the Galactic center.}
\label{fig:halo} 
\end{figure}

Since it is possible relic WIMPs are still annihilating in our Galactic halo, 
the ID detection rates
mentioned in Sec.~\ref{sec:intro} depend on the assumed galactic DM density (halo)
profile. We show several popular halo profiles in
Fig. \ref{fig:halo}. Most models are in near accord at the earth's
position at $\sim 8$ kpc from the galactic center. However, we see that
predictions for the DM density near the galactic center differ wildly,
which translates to large uncertainties for DM annihilation rates
near the galactic core. The corresponding uncertainty
will be smaller for anti-protons, and smaller still for positrons, since
these particles gradually lose energy while propagating through the
galaxy, and so can reach us from limited distances over which
the halo density is relatively well-known.
Possible clumping of DM yields an additional source of uncertainty in ID detection rates. 

DD rates are determined by the local DM density (usually taken
to be $\rho_{local}\simeq 0.3$ GeV/cm$^3$), the WIMP mass and
the WIMP-nucleon scattering cross section. Most experiments are sensitive
mainly to the {\it spin-independent} WIMP-nucleon cross section, since
in this case WIMP scattering rates are $\propto A^2$ (where $A$ is the
mass number of the nuclear target) because the WIMP here couples {\it
coherently} to the entire nucleus: hence its scattering cross section is 
amplified for heavy nuclei.

\begin{figure}[tbh]
\begin{center}
\includegraphics[width=12cm,angle=-90]{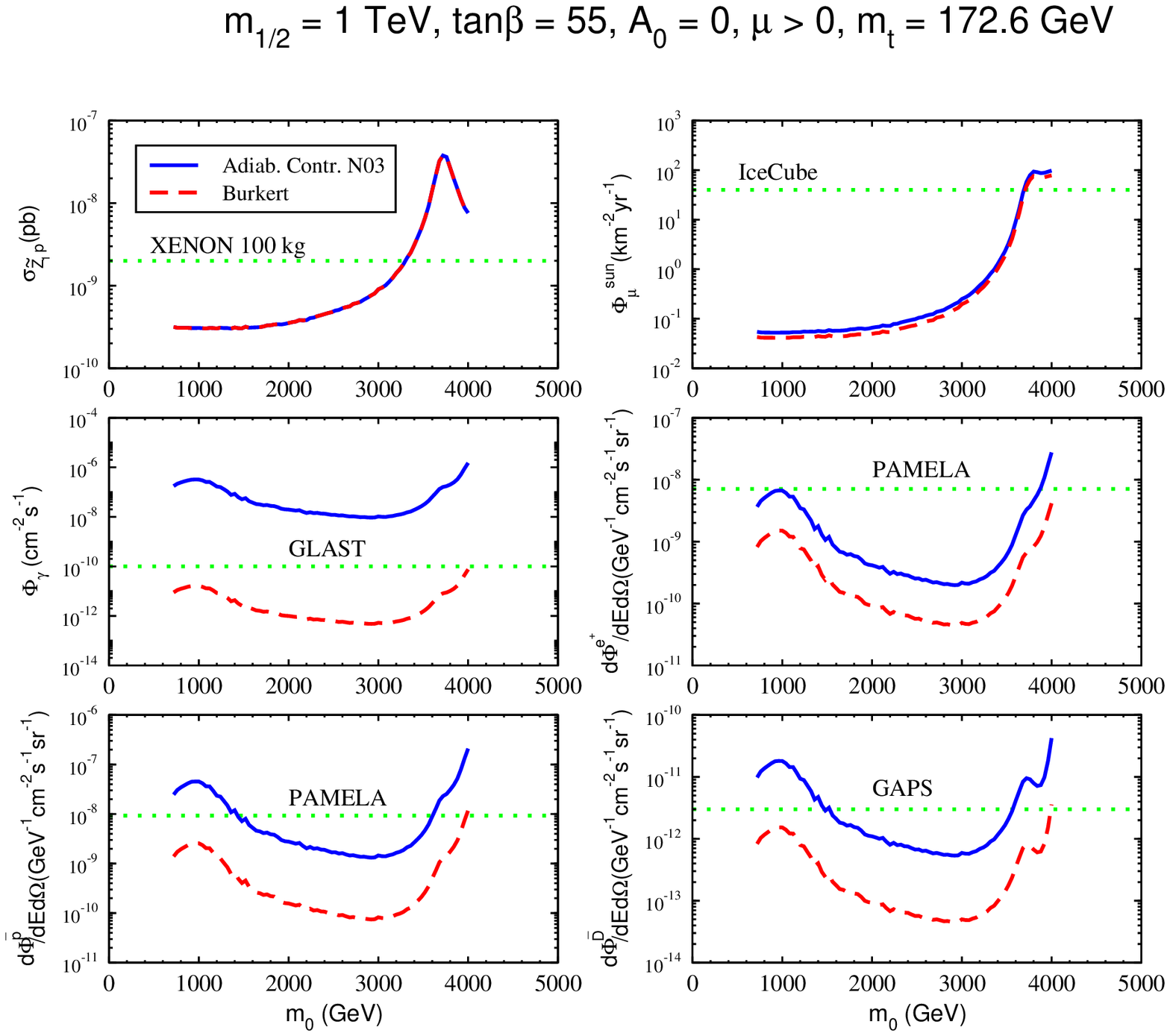}
\end{center}
\vspace{-3mm}
\caption{\small\it Projected rates for direct and indirect detection of
neutralino dark matter in mSUGRA model for the Burkert and Adiabatically
contracted N03 halo profiles, with
mSUGRA parameters as listed. Also shown are expected sensitivities of
various experiments. }
\label{fig:dd_idd} 
\end{figure}

We have calculated the cross section $\sigma_{\rm SI}(\tz_1 p)$ via the
scalar interaction using the program IsaReS\cite{isares}.  We show some
results from the mSUGRA model in Fig. \ref{fig:dd_idd}{\it a}), where we
fix mSUGRA parameters $m_{1/2}=1$ TeV, $A_0=0$ and $\tan\beta =55$.  We
plot the cross section against variation in $m_0$.  At low $m_0\sim700$
GeV, we are in the stau co-annihilation region, and the $\tz_1$ is
nearly bino-like. Here, the DD rates are well below the projected
sensitivity of the Xenon-100 or LUX experiments, depicted by the dotted
line, which shows the sensitivity for a 100~GeV neutralino. (For a
bino-like neutralino with a mass $\sim 400$~GeV that obtains for
$m_0 \alt 2$~TeV, the detectability level is about twice this.) 
As $m_0$ steadily increases, $m_{\tz_1}$ changes only slowly until
the magnitude of the $\mu$ parameter drops to sufficiently low values
and the $\tz_1$ becomes increasingly higgsino-like.  The $\tz_1$
coupling to Higgs bosons increases, as does $\sigma_{\rm SI}(\tz_1 p )$.
In the HB/FP region, the cross section reaches above the $10^{-8}$ pb
level, within the reach of the next round of experiments.

In Fig. \ref{fig:dd_idd}{\it b}), we show the flux of muons from
$\nu_\mu\to \mu$ conversions at earth coming from neutralino
annihilation to SM particles within the solar core.  Here, we use the
Isajet/DarkSUSY interface for our calculations\cite{dsusy}, and require
$E_\mu >50$ GeV.  The predicted rate depends in this case mainly on the
sun's ability to sweep up and capture neutralinos, which depends mainly
on the {\it spin-dependent} neutralino-nucleon scattering cross section
(since in this case, the neutralinos mainly scatter from solar Hydrogen,
and there is no mass number enhancement), mostly sensitive to $Z$
exchange. The rates are again low for low $m_0$ with bino-like
neutralinos, but reach the IceCube detectability level at large $m_0$ in
the HB/FP region where neutralino couplings to $Z$ become large. 

In Fig. \ref{fig:dd_idd}{\it c}), we show the expected flux of gamma
rays with $E_\gamma >1$ GeV, as required for the Fermi Gamma-ray Space
Telescope (FGST), arising from DM annihilations in the galactic core. In
this case, we see enhanced signal at both low $m_0$ and high
$m_0$. The low $m_0$ enhancement occurs because we are at high
$\tan\beta =55$, and neutralinos can annihilate efficiently through the
$A$-resonance since here $2m_{\tz_1}\sim m_A$\cite{bo}. 
As we move to higher $m_0$, $m_A$
increases, and we move out of the $A$ funnel.  At very large $m_0$, we
are back to the HB/FP region, and $\tz_1\tz_1\to WW$, $ZZ$ and
$t\bar{t}$ are all enhanced, and we get elevated gamma ray detection
rates.  The predictions for two halo profiles differ by four
orders of magnitude, reflecting the large uncertainty in our knowledge
of the DM density at the center of our Galaxy.

In Fig. \ref{fig:dd_idd}{\it d})-{\it f})., we show the expected flux of
positrons, $\bar{p}$s and antideuterons $\overline{D}$ from neutralino
halo annihilations.\footnote{Several groups \cite{pulsars} have recently
noted that positrons (but not anti-protons or anti-deuterons) with
energies up to ${\cal O}(100)$~GeV can be produced in local pulsars. It
will be essential to understand the level of this pulsar background to
any positron signal from annihilating DM. It would be also interesting
to study whether collisions of protons, accelerated by the same
mechanism as positrons, with matter in the environment can produce a
detectable flux of high energy neutrinos pointing back to the
pulsar.} Each of these frames show elevated rates in the $A$-funnel and
in the HB/FP region. 
The various rates shown in this figure exemplify the possibility of a
discrimination between DM annihilation mechanisms in the early
universe\cite{bo}. If we are in the stau co-annihilation region, we
expect very low rates for both DD and ID experiments, possibly  with 
characteristic implications for the LHC \cite{aandm}.  In the
$A$-funnel, we expect low rates for DD and ID via $\nu_\mu$ telescopes,
but enhanced rates for ID via gamma and anti-matter searches.  If we are
in the HB/FP region, then DD, ID via muons and ID via halo annihilations
would all expect to be elevated, and possibly observable.

\subsection{Dark matter at colliders: reach plots}

\begin{figure}[tbh]
\begin{center}
\includegraphics[width=7.5cm]{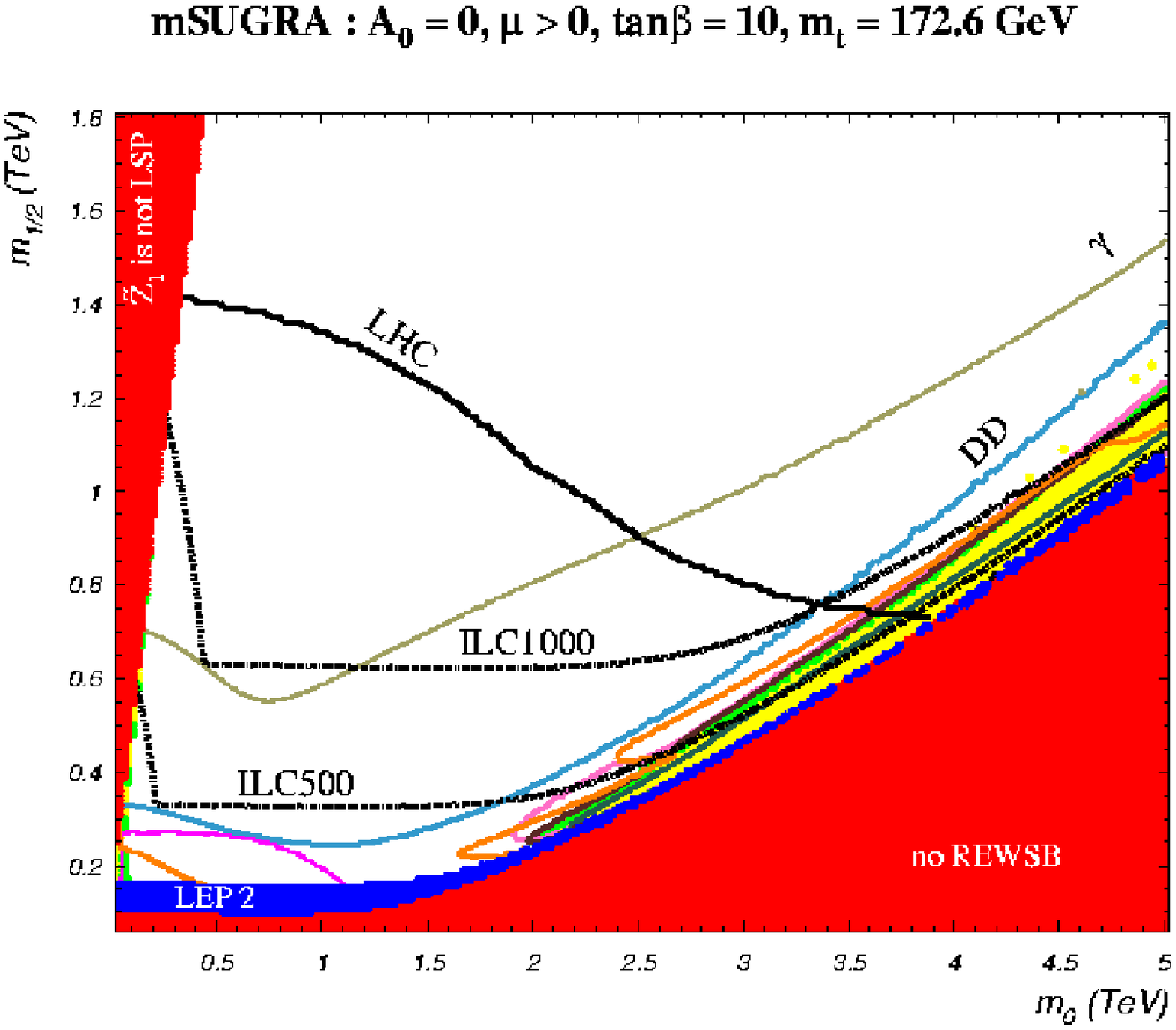}
\includegraphics[width=7.5cm]{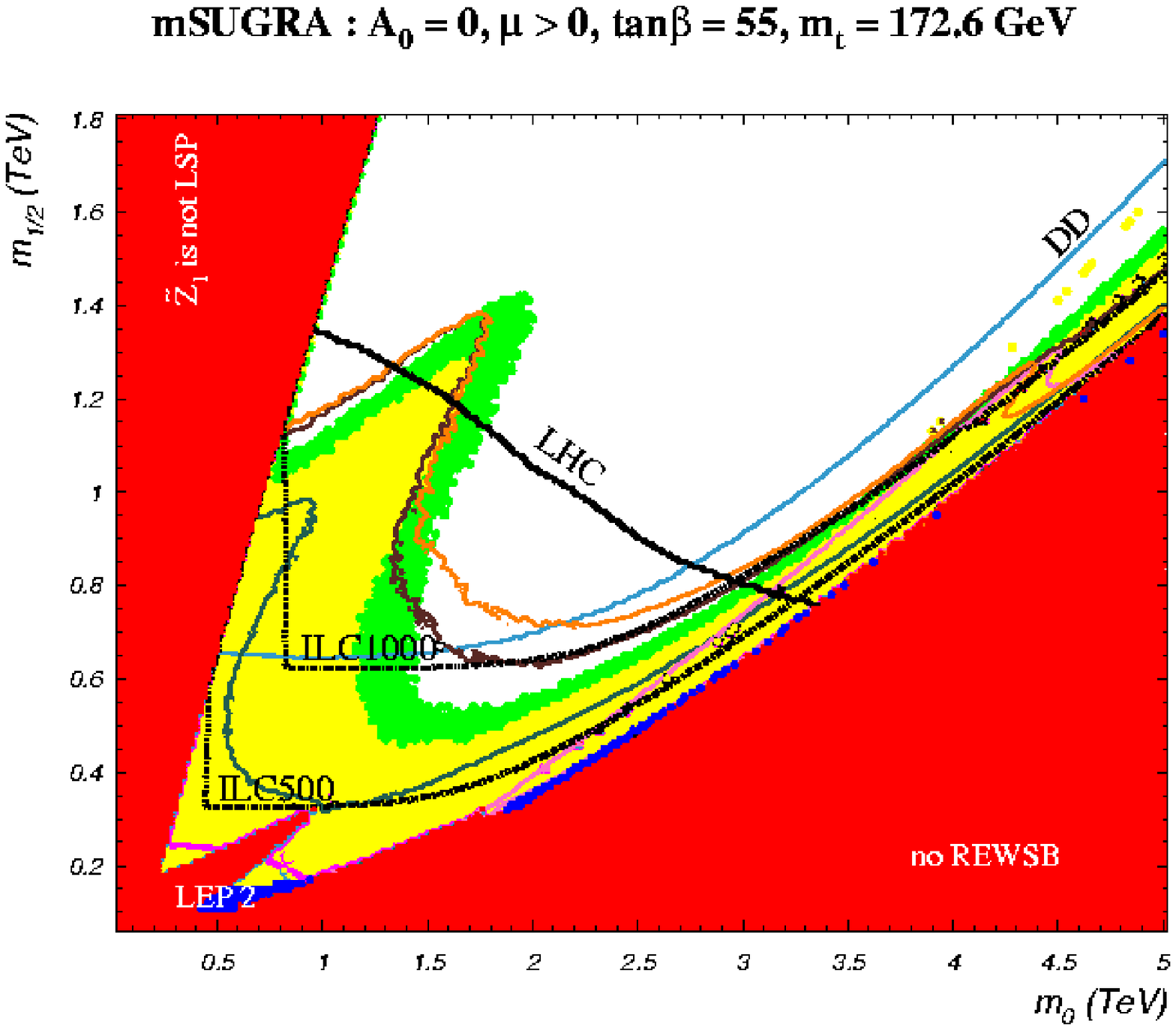}
\end{center}
\vspace{-10mm}
\begin{center}
\includegraphics[width=10cm]{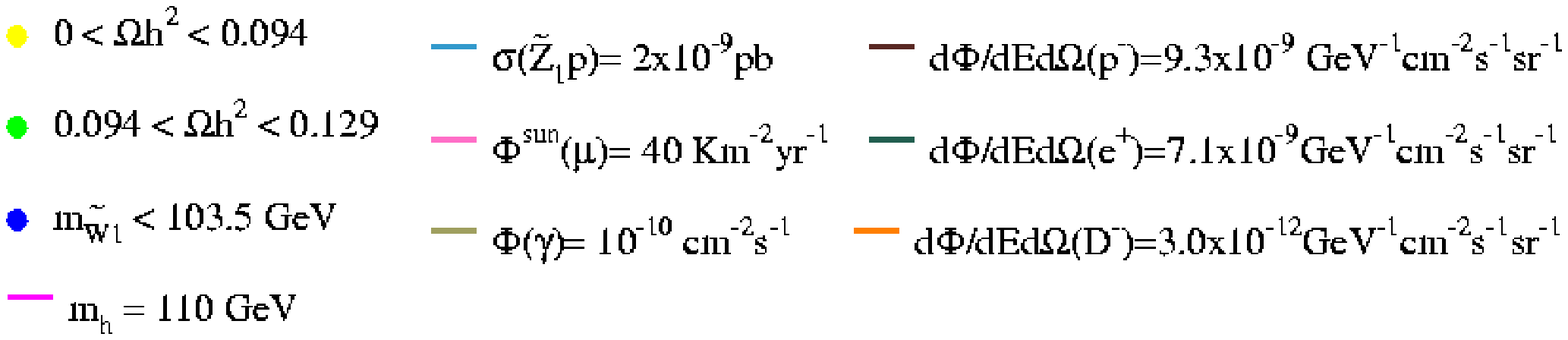}
\end{center}
\vspace{-3mm}
\caption{\small\it The projected reach of various colliders, direct and
  indirect dark matter search experiments in the $m_0\ vs.\ m_{1/2}$
  plane of the mSUGRA model for $A_0=0$, $\mu >0$, $m_t=172.6$ GeV for
  $\tan\beta =10$ (left frame) and $\tan\beta =55$ (right frame). The DD
  and various ID contours are for the corresponding expected
  sensitivity in Fig.~\ref{fig:dd_idd}.  For the ID results, we
  have adopted the N03 DM halo density profile.}
\label{fig:pm10} 
\end{figure}

In Fig.~\ref{fig:pm10}, we show the SUSY reach of various experiments in
the $m_0-m_{1/2}$ plane of the mSUGRA model for a low (left frame) and
high (right frame) value of $\tan\beta$.  The approximate SUSY reach of
the LHC, assuming an integrated luminosity of 100~fb$^{-1}$, and of the
proposed $e^+e^-$ International Linear Collider operating at
$\sqrt{s}=0.5$ or 1~TeV are depicted by the correspondingly labelled
contours.  The LHC reach contour is a {\it cumulative} contour, but the
largest reach appears in the inclusive multi-jet $+\eslt$
channel\cite{lhcreach}. In much of the accessible parameter space,
signals in several different event topologies with differing numbers of hard,
isolated leptons should be visible as well.  This will help add
confidence that one is actually seeing new physics, and may help to sort
out the production and decay mechanisms.  The reach at low $m_0$ extends
to $m_{1/2}\sim 1400$~GeV. This corresponds to a reach for $m_{\tq}\sim
m_{\tg}\sim 3.1$ TeV.  At large $m_0$, squarks and sleptons are in the
$4-5$ TeV range, and are too heavy to be produced at significant rates
at LHC. Here, the reach comes mainly from just gluino pair
production. In this range, the LHC reach is up to $m_{1/2}\sim 700$ GeV,
corresponding to a reach in $m_{\tg}$ of about 1.8 TeV, and may be
extended by $\sim$ 15-20\% by $b$-jet tagging\cite{mizukoshi}.
%
While LHC can cover  the relic density allowed bulk and
stau co-annihilation regions, as well as most of 
the $A$-funnel region that appears only for large $\tan\beta$, the HB/FP
region extends far beyond the LHC reach. 
The ILC(1000) reach is everywhere lower than LHC, except in the HB/FP
region.  In this region, while gluinos and squarks can be extremely
heavy, the $\mu$ parameter is small, leading to a relatively light
spectrum of charginos and neutralinos. These are not detectable at the
LHC because the visible decay products are too soft.  However, with
specialized cuts, 
chargino pair production is detectable at ILC even if the energy release
in chargino decays is small, and the ILC reach extends beyond LHC in this
region\cite{ilcreach}.

In Fig.~\ref{fig:pm10}, we also show reach contours for DD and ID
searches for WIMP dark matter\cite{bo}. Signals from DD are observable in 
{\it i}). the region of low $m_0$ and low $m_{1/2}$, where squarks are light and
scattering via squark exchange occurs, and {\it ii}). also in
the {\it entire} HB/FP region (where $\tz_1$ is MHDM) where the reach of the LHC
is limited to $m_{1/2}\alt 700$ GeV. 
Thus, in the HB/FP region with $m_{1/2}>700$ GeV, it is possible a DM direct
detection signal might be seen, while no signal is evident from LHC.
The DD rate increases with $\tan\beta$, accounting for the
shift in the corresponding contour in the right hand frame. 

The $\nu_\mu$ rates at IceCube/Antares are largest in the HB/FP region,
where spin-dependent scattering cross sections are large.  For the
peaked N03 halo profile used here, the $\gamma$ signal is observable
over a large part (the entire) plane in the left (right) frame, though
we caution that this is very sensitive to the assumed profile. The
expected $\bar{p}$ and $\overline{D}$ signals are large in the HB/FP
region, and also cover much of the $A$-funnel in the right-hand frame,
while positron signals are observable over a smaller region. For $\mu <
0$ and large $\tan\beta$, $A$ is lighter than for $\mu >0$, and the $A$-funnel
extends well beyond the reach of the LHC; again, for this 
halo profile, ID anti-particle signals cover much of the $A$-funnel region. 
%
%

%
%

\subsection{Characterizing dark matter at collider experiments}

SUSY discovery will undoubtedly be followed by a program to reconstruct
sparticle masses, couplings and quantum numbers.  What will we be able
to say about dark matter in light of these measurements?  Several groups
have made such studies \cite{bbpw}.  Baltz {\it et al.} examined four
mSUGRA case study points (one each in the bulk region, the HB/FP region,
the stau co-anihilation region and the $A$-funnel region). They extract from
other studies the precision with which various sparticle
properties could be measured at LHC, and also at a $\sqrt{s}=0.5$ and 1
TeV $e^+e^-$ collider.  They then adopted a 24-parameter version of the
MSSM, fit its parameters to these projected measurements, and used the result
to predict several quantities relevant to astrophysics and
cosmology: the dark matter relic density $\Omega_{\tz_1}h^2$, the
spin-independent neutralino-nucleon scattering cross section
$\sigma_{SI}(\tz_1 p)$, and the neutralino annihilation cross section
times relative velocity, in the limit that $v\to 0$: $\langle\sigma
v\rangle |_{v\to 0}$.  This last quantity is the crucial particle physics
input for estimating signal strength from neutralino annihilation to
anti-matter or gammas in the galactic halo.  What this yields then is a
{\it collider measurement} of these key dark matter quantities. Arnowitt
{\it et al.} \cite{aandm} performed detailed studies of mSUGRA points 
in the stau co-annihilation region to project the precision with which LHC
can ``measure'' the neutralino relic density.

\begin{figure}[tbh]
\begin{center}
\includegraphics[width=7cm]{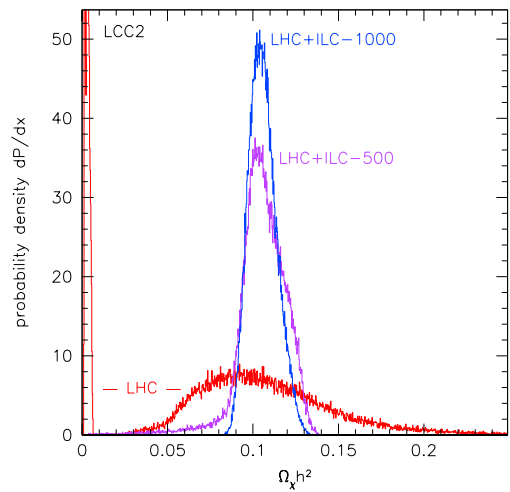}
\end{center}
\vspace{-3mm} 
\caption{\small\it 
Determination  of neutralino relic abundance via measurements
at the LHC and ILC, taken  from Baltz {\it et al.}~\cite{bbpw}.}
\label{fig:lcc2} 
\end{figure}

As an illustration, we show in Fig. \ref{fig:lcc2} (taken from Baltz
{\it et al.} \cite{bbpw}) the precision with which the neutralino relic
density is constrained by collider measurements for the LCC2 point which
is in the HB/FP region of the mSUGRA model. Measurements at the LHC
cannot fix the $\tz_1$ composition, and so are unable to resolve the
degeneracy between a wino-LSP solution (which gives a tiny relic
density) and the true solution with MHDM. Determinations of chargino
production cross sections at the ILC can easily resolve the
difference. It is nonetheless striking that up to this degeneracy
ambiguity, experiments at the LHC can pin down the relic density to
within $\sim 50$\% (a remarkable result, given that there are sensible
models where the predicted relic density may differ by orders of
magnitude!). This improves to 10-20\% if we can combine LHC and ILC
measurements.

A collider determination of the relic density is very important. If
it agrees with the cosmological measurement it would establish that the
DM is dominantly thermal neutralinos from the Big Bang.  If the
neutralino relic density from colliders falls significantly below the
measured CDM density, it would provide direct evidence for
multi-component DM-- perhaps neutralinos plus axions or other
exotica. Alternatively, if the collider determination gives a much
larger value of $\Omega_{\tz_1}h^2$, it could point to a long-lived but
unstable neutralino and/or non-thermal DM.

The collider determination of model parameters would also pin down the
neutralino-nucleon scattering cross section. Then if a WIMP signal is
actually observed in DD experiments, one might be able to determine the
local DM density of neutralinos and aspects of their velocity
distribution based on the DD signal rate. This density should agree with
that obtained from astrophysics if the DM in our Galaxy is comprised
only of neutralinos.

Finally, a collider determination of $\langle\sigma v\rangle |_{v\to 0}$
would eliminate uncertainty on the particle physics side of projections for
any ID signal from annihilation of neutralinos in the galactic halo.
Thus, the observation of a gamma ray and/or anti-matter signal from 
neutralino halo annihilations would facilitate the determination
of the galactic dark matter density profile.

\subsection{Non-universal SUGRA models: the well-tempered neutralino}

\begin{figure}[hbt]
\begin{center}
\includegraphics[width=10cm]{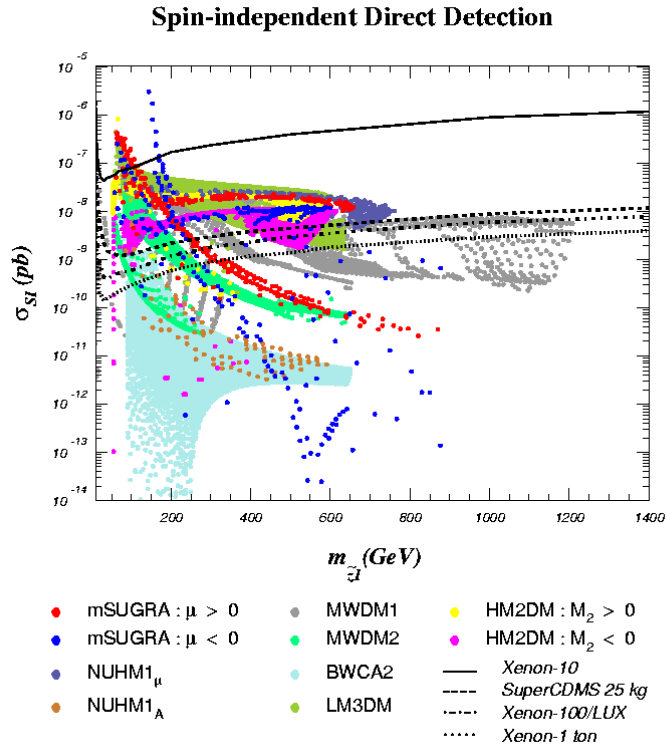}\\
\end{center}
\vspace{-3mm}
\caption{\small\it 
The spin-independent neutralino-proton scattering cross-section vs
$m_{\tz_1}$ in a variety of SUSY models, compatible
with collider constraints where thermally produced  Big Bang neutralinos
saturate the observed dark matter density.}
\label{fig:dd} 
\end{figure}

The underlying universality of scalar mSUGRA parameters results from a
technical assumption and so has a rather weak theoretical motivation
\cite{soni}. Unification of gaugino mass parameters may also not obtain even in
SUSY GUT models if the order parameter for SUSY breaking also breaks the
GUT symmetry \cite{nonunigaug}. It is, therefore, of interest to
consider models with non-universal SSB parameters.

In Fig. \ref{fig:dd}, we show the spin-independent $\tz_1 p$ cross
section versus $m_{\tz_1}$ for a large number of one-parameter
extensions of mSUGRA, where the GUT scale universality between matter
scalar and Higgs scalar mass parameters, or between the three gaugino
mass parameters is relaxed in a systematic way.  The details of the
various models are not essential for our present purpose, but may be
found in Ref.~\cite{wtnreview}.  In each such model, shown by a
different colour on the plot, this additional parameter is adjusted so
that the lightest neutralino (assumed to be the LSP) {\it saturates} the
observed relic abundance of CDM.  We also include the mSUGRA model. To
make this plot, we randomly generated points in the parameter space for
each model, and plotted it on the figure if all current collider
constraints on sparticle masses are satisfied. We also show the
sensitivity of current experiments together with projected sensitivity
of proposed searches at superCDMS, Xenon-100, LUX, WARP and at a
ton-sized noble liquid detector.  The key feature to note is that while
the various models have a branch where $\sigma_{\rm SI}(p\tz_1)$ falls
off with $m_{\tz_1}$, there is another branch where this cross-section
asymptotes to $\sim 10^{-8}$~pb\cite{wtn,wtnreview,pran}.  This branch
(which includes the HB/FP region of mSUGRA) includes {\it many} models
with MHDM which easily accommodate the measured relic density via {\it
tempering} of the neutralino's higgsino content.  In these cases, the
spin-independent DD amplitude -- which is mostly determined by the Higgs
boson-higgsino-gaugino coupling -- is large because the neutralino has
both gaugino and higgsino components.
The exciting thing is that the experiments currently being deployed--
such as Xenon-100, LUX, WARP and superCDMS -- will have the 
necessary sensitivity to probe this {\it entire class of models}! 
To go further will require ton-size or larger detectors.

We note here that if $m_{\rm WIMP}\alt 150$ GeV, then it may be possible
to extract the WIMP mass by measuring the energy spectrum of the
recoiling nuclear targets\cite{green}.  Typically, of order 100 or more
events are needed for such a determination to 10-20\%. For higher WIMP
masses, the recoil energy spectrum varies little, and WIMP mass
extraction is much more difficult.  Since the energy transfer from the
WIMP to a nucleus is maximized when the two have the same mass, DD
experiments with several target nuclei with a wide range of
masses would facilitate the distinction between somewhat light and
relatively heavy WIMPs, and so potentially serve to establish the
existence of multiple WIMP components in our halo.

Before closing this section, we remark that in the various one-parameter
extensions of mSUGRA that we have considered, {\it any point} in the
$m_0-m_{1/2}$ plane can be made consistent with the measured relic
density. We therefore caution drawing inferences about collider signals
from the relic density measurement from any analysis based on just the
mSUGRA framework. Based on the analysis of the various one-parameter
extensions of mSUGRA that we have studied \cite{wtnreview}, we infer
that in most relic-density-consistent models: 1)~$m_{\tq}\sim m_{\tg}$
so that the LHC reach extends to about $m_{\tg}\sim 3$~TeV,
2)~$m_{\tz_2}-m_{\tz_1} < M_Z$, so that there should be a discernable
edge in the opposite-sign, same-flavour dilepton mass distribution in
SUSY events that can serve as a starting point for sparticle mass
reconstruction at the LHC, 3)~the mechanism that increases the
neutralino annihilation rate frequently also enhances the direct or
indirect rates for DM searches. In this connection, we remark that
inclusion of neutrino Yukawa couplings as given by a $SO(10)$ SUSY GUT
see-saw, significantly changes the location of the
relic-density-consistent region in SUSY parameter space, but has little
impact on the DD and ID detection rates \cite{bmm}.

\section{WIMP signals in cosmic ray data?} \label{sec:indirect}

Various indirect searches for DM have {\it already} turned up
suggestive hints of a possible WIMP signal.
These include:
\bi
\item The HEAT experiment, in balloon flights from 1994, 1995 and 2000,
measured an excess of positrons in cosmic ray data with energies in the
range 10-30 GeV.  Their measured rate is above that expected from WIMP
dark matter annihilations, unless a substantial ``boost'' factor
(enhancement due to fluctuations in the dark matter density
distribution) of order 50 is included in the theoretical
projections\cite{heat}. However, it now seems likely that they are seeing
the influence of cosmic {\it protons}-- rather than positrons--
which actually ought to manifest themselves at high energy.

\item The EGRET GeV anomaly: Here, a detection of an excess of around
  0.5-5~GeV $\gamma$ rays above background projections has been
  interpreted as possible WIMP annihilation into $b\bar{b}$
  states\cite{willem}.  This interpretation requires ring-like
  structures in the Miky Way DM density profile, along with galactic
  magnetic fields that sweep anti-protons out of the galactic disk.
  Although an interpretation\cite{willem2} in terms of the mSUGRA model
  seems to contradict DD search limits from Xenon-10 and CDMS2, assuming
  a standard local DM density, the data can be accommodated by models
  with non-universal Higgs mass parameters \cite{bbs}.  However, it
  recently appears that the latest FGST data are {\it in accord} with
  background expectations\cite{mosk}, which may end up ruling out this
  galactic EGRET anomaly.

\item The multi-GeV extra-galactic gamma ray anomaly, suggested by the
EGRET observation of an apparent excess 1-20 GeV gamma rays has been
interpreted as annihilation of a 500 GeV WIMP, and requires cuspy DM
profiles in other galaxies which are not seen in the Milky Way\cite{egret2}.

\item The WMAP collaboration measures an excess of microwave emissions
from the galactic core. 
It has been suggested that WIMP annihilation in
the galactic core into $e^+e^-$ pairs, with subsequent synchroton
emissions, could explain this WMAP Haze\cite{fink}

\item Several particle physics explanations have been suggested to
account for the excess of positrons with $E_{e^+}\sim 10-100$ GeV
claimed by the Pamela collaboration \cite{pamela}, and the excess of
electrons and/or positrons with $E_{e^\pm}\sim 300-800$~GeV claimed by
the ATIC balloon experiment \cite{atic}.  The explanations, which do not
accommodate the possible structure seen in the ATIC energy spectrum, are
also constrained by the fact that the measured $\bar{p}$ flux is
consistent with SM predictions. Questions have also been raised as to
just how well Pamela can discriminate {\it protons} from $e^+$s. Also, as
noted in an earlier footnote, it appears possible to accommodate the claimed
Pamela/ATIC signals in terms of acceleration of positrons produced via
$\gamma\gamma\to e^+e^-$ in nearby pulsars \cite{pulsars}; this explanation
 naturally accounts for the non-observation of an excess of high energy
 anti-protons. 

\ei

\section{Gravitinos} \label{sec:gravitino}

\subsection{The gravitino problem}

\begin{figure}[tbh]
\begin{center}
\includegraphics[width=7cm]{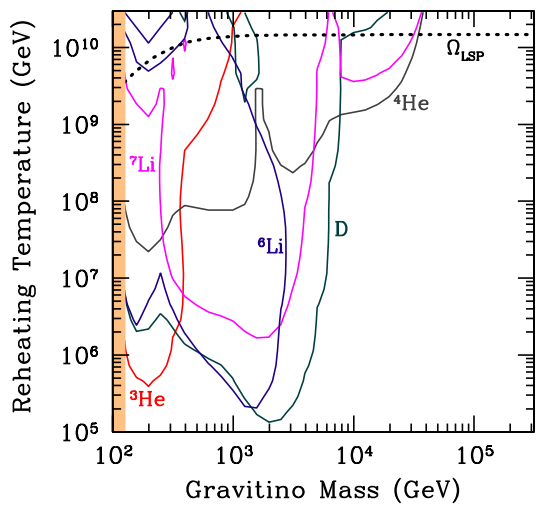}
\end{center}
\vspace{-3mm}
\caption{\small\it An illustration of constraints from Big Bang
nucleosynthesis which require $T_R$ to be below the various curves, for the
HB/FP region of the mSUGRA model with $m_0=2397$ GeV, $m_{1/2}=300$ GeV,
$A_0=0$ and $\tan\beta=30$, from  Ref. \cite{moroi2}, to
which we refer the reader for more details.}
\label{fig:gino} 
\end{figure}

In gravity-mediated SUSY breaking models, gravitinos typically have weak
scale masses and, because they only have tiny gravitational couplings,
are usually assumed to be irrelevant for particle physics phenomenology.
Despite their tiny coupling, they are not irrelevant for cosmology where
we may have the {\it gravitino problem}. Though not the LSP, 
gravitinos -- while
not in thermal equilibrium-- can be 
produced in the early universe via
emission from particles that are in thermal equilibrium. 
These {\it thermally produced}  gravitinos
then decay with a lifetime which is very roughly $\tau \sim
G/m_{\tG}^3$, typically well after
Big Bang nucleosynthesis (BBN).  The late-time injection of hadronic
(and electromagnetic) energy from these gravitino decays into the cosmic
soup can again disrupt the successful predictions of BBN
\cite{gravitinop,moroi2}.  The precise constraints depend on the
gravitino mass, the re-heating temperature $T_R$ of the universe after
inflation, and to a smaller extent on the various sparticle masses and
mixings.  We illustrate this in Fig.~\ref{fig:gino} where the constraint
on the gravitino mass is shown as a function of the re-heating
temperature for a case study in the HB/FP region of the mSUGRA model. We
see that it is possible to accommodate $m_{\tG}\alt 3$ TeV and avoid
disruption of BBN if $T_R\alt 10^5$ GeV. Such a low re-heat temperature
puts severe constraints on inflationary models, and also call for rather
low temperature baryogenesis mechanisms\cite{lowscale}.

Alternatively, we may assume that the massive 
$\tG$ is in fact the stable LSP, and thus constitutes the DM
\cite{primack,feng}.
In this case, one has to worry about thermal production of SUSY particles, 
followed by their late time decay in SM particles plus the gravitino
since this may again disrupt the successful BBN predictions.

Finally, we remark here upon the interesting interplay of baryogenesis
via leptogenesis with the nature of the LSP and the
next-lightest-supersymmetric-particle (NLSP). For successful thermal
leptogenesis to take place, it is found that the reheat temperature of
the universe must exceed $\sim 10^{9}$ GeV\cite{buchmuller}.  If this is
so, then gravitinos would be produced thermally with a huge abundance,
and then decay late, destroying BBN predictions. For this reason, some
adherents of leptogenesis tend to favor scenarios with a gravitino LSP,
but with a stau NLSP\cite{buchm}. A recent study \cite{moroi2} suggests
that the gravitino LSP then has to be lighter than about 10~GeV unless
$m_{\ttau} > 1$~TeV, implying a very heavy sparticle spectrum.

\subsection{Gravitinos as dark matter}

Here, we consider the consequences of a gravitino LSP in SUGRA 
models\cite{feng,gmodels}. 
If gravitinos are produced in the pre-inflation epoch, then
their number density will be diluted away during inflation. After the
universe inflates, it enters a re-heating period wherein all particles
can be thermally produced. However, the couplings of the gravitino are
so weak that though gravitinos can be produced by particles in 
thermal equilibrium, gravitinos themselves never attain
thermal equilibrium: indeed their density is so low that gravitino
annihilation processes can be neglected in the calculation of their
relic density.  The thermal production (TP) of gravitinos in the early
universe has been calculated and, including EW contributions, is given
by the approximate expression (valid for $m_{\tG}\ll M_i$\cite{thermal_G}):
\be
\Omega_{\tG}^{TP}h^2\simeq 0.32\left(\frac{10\ GeV}{m_{\tG}}\right)
\left(\frac{m_{1/2}}{1\ {\rm TeV}}\right)^2\left(\frac{T_R}{10^8\
  {\rm GeV}}\right)\; 
\ee
where $T_R$ is the re-heat temperature.
%
%

If gravitinos are the LSP, then they can also be produced by decay of
the NLSP.  In the case of a long-lived neutralino NLSP, the neutralinos
will be produced as usual with a thermal relic abundance in the early
universe. They will subsequently decay via $\tz_1\to \gamma \tG,\ Z\tG$ or
$h\tG$. Thus, the non-thermally produced gravitinos inherit the
thermally produced neutralino number density. The total relic abundance
is then
\be
\Omega_{\tG}h^2 =\Omega_{\tG}^{TP}h^2+\frac{m_{\tG}}{m_{\tz_1}}\Omega_{\tz_1}h^2.
\ee
The $\tG$ from NLSP decay may constitute warm/hot dark matter depending
in the $\tz_1 -\tG$ mass gap, while the thermally produced $\tG$ will be
cold DM\cite{jlm}.

The lifetime for neutralino decay to photon plus gravitino is
given by \cite{fst}, 
\bea
\hspace{-5mm}\tau (\tz_1\to \gamma\tG ) \simeq {{48\pi M_P^2}\over
  m_{\tz_1}^3} A^2 {{r^2}\over{(1-r^2)^3(1+3r^2)}} \nonumber \\
\hspace{1mm}\sim 5.8\times 10^8 \ {\rm s} \left({100 \ {\rm GeV}}\over
{m_{\tz_1}}\right)^3
{1\over
  A^2} {{r^2}\over{(1-r^2)^3(1+3r^2)}}\;, 
\label{gravdec}
\eea where $A=(v_4^{(1)}\cos\theta_W+v_3^{(1)}\sin\theta_W)^{-1}$, with
$v_{3,4}^{(1)}$ being the wino and bino components of the $\tz_1$, in
the notation of the first item of Ref.~\cite{wss}, $M_P$ is the reduced
Planck mass, and $r=m_{\tG}/m_{\tz_1}$. Similar formulae (with different
mixing angle and $r$-dependence) hold for decays to the gravitino plus a
$Z$ or $h$ boson.  We see that -- except when the gravitino is very much
lighter than the neutralino as may be the case in GMSB models with a low
SUSY breaking scale -- the NLSP decays well after Big Bang
nucleosynthesis.  Such decays would inject high energy gammas and/or
hadrons into the cosmic soup post-nucleosynthesis, which could break up
the nuclei, thus conflicting with the successful BBN predictions of Big
Bang cosmology.  For this reason, the gravitino LSP scenarios usually
favor a stau NLSP, since the BBN constraints in this case are much
weaker: see, however, Ref.~\cite{moroi2} where it is noted that bounds
from $^6$Li abundance constrain the gravitino to be lighter than 10~GeV
unless the stau is heavier than 1~TeV. 

Before closing this section, we remark that the NLSP could be
electrically charged or coloured. It will then be revealed via
specialized searches for for quasi-stable, slow-moving particles
\cite{slow}. More strikingly, it may be possible to trap the
very-long-lived ($\tau \sim 10^5-10^8$~s) NLSPs produced at high energy
colliders, and then search for their subsequent decays \cite{feng2}.

\section{Mixed axion/axino dark matter} \label{sec:axion}

\subsection{Axion dark matter}

The axion arises as a by-product of the Peccei-Quinn solution to the
strong $CP$ problem\cite{pqww,axreview}. The strong $CP$ problem has its
origin in an allowed QCD Lagrangian term,
\be
{\cal L}\ni \frac{\theta g_s^2}{32\pi^2}G_{\mu\nu}^a\tilde{G}^{a\mu\nu}\;,
\ee
(here, $G_{\mu\nu}^a$is the gluon field strength tensor and
$\tilde{G}^{a\mu\nu}$ its dual) which is both $P$- and $T$-violating, and
hence $CP$-violating.  When QCD is coupled to the electroweak theory,
$\theta$ is replaced by $\bar{\theta}\equiv \theta +arg(det\ m_q )$,
where $m_q$ is the quark mass matrix.  The measured value of the neutron
electric dipole moment (EDM) requires $\bar{\theta}\alt 10^{-10}$.
Explaining the tininess of this Lagrangian term is the strong $CP$
problem.

The Peccei-Quinn solution to the strong $CP$ problem promotes $\theta$
to a field in a theory with a global $U(1)$ (Peccei-Quinn or PQ)
symmetry, that is broken spontaneously, and by instanton effects. A
consequence of the broken PQ symmetry is the existence of a
pseudo-Goldstone boson field: the axion $a(x)$\cite{pqww}, which
acquires a small mass due to instanton effects.  The axion Lagrangian
includes the terms
\be
{\cal L}\ni {1\over 2}\partial_\mu a\partial^\mu a +\frac{g^2}{32\pi^2}\frac{a(x)}{f_a/N}
G_{\mu\nu}^a \tilde{G}^{a\mu\nu} , \label{L_ax}
\ee
where we have introduced the PQ breaking scale $f_a$ and $N$ is the
model-dependent color anomaly of order 1.  The effective potential for
the axion field $V(a(x))$ has its minimum at $\langle a(x)\rangle =
-\bar{\theta} f_a/N$, and so the offending $G\tilde{G}$ term essentially
vanishes, solving the strong $CP$ problem.  An inescapable consequence 
of this mechanism is the existence of axions
with a mass
given by,
\be m_a\simeq 6\ {\rm eV}\frac{10^6\ {\rm GeV}}{f_a/N}\;, 
\label{eq:axmass}
\ee
and coupled to gluons as in (\ref{L_ax}), and to photons by an analogous
term, with the coupling constant suppressed by the PQ scale, $f_a$. 


Astrophysical limits from cooling of red giant stars and supernova 1987a
require $f_a/N \agt 10^9$ GeV, or $m_a\alt 3\times 10^{-3}$ eV. In
addition, axions can be produced via various mechanisms in the early
universe. Since their lifetime (they decay via $a\to\gamma\gamma$) turns
out to be longer than the age of the universe, they can be a good
candidate for dark matter in the universe.  In SUGRA models, we will be
concerned with re-heat temperatures of the universe $T_R\alt 10^9\ {\rm
GeV}<f_a$ (to avoid overproducing gravitinos in the early universe), the
axion production mechanism relevant for us here is just one: production
via vacuum mis-alignment\cite{absik}. In this mechanism, the axion field
$a(x)$ is homogenized by inflation (assumed to occur after the PQ
phase transition), and can have any value
$\alt f_a$ at temperatures $T\gg \Lambda_{QCD}$. As the
temperature of the universe drops to the quark-hadron phase transition
temperature, the axion potential turns on, and the axion field
oscillates about its minimum at $-\bar{\theta} f_a/N$, resulting in the
production of {\it non-relativistic} axions from the nearly homogeneous
condensate.  This
``vacuum mis-alignment'' mechanism for axion production thus results in
  {\it cold axion dark matter} with a number density,
\be
n_a(t)\sim {1\over 2}m_a(t)\langle a^2(t)\rangle ,
\ee
where $t$ is  the time near the QCD phase transition.
Relating the number density to the entropy density allows one to determine the
axion relic density today to be
\be
\Omega_a h^2\simeq {1\over 4}\left(\frac{6\times 10^{-6}\ {\rm eV}}{m_a}\right)^{7/6}\;,
\label{eq:axrelic}
\ee
to within a factor of about three.

\begin{figure}[tbh]
\begin{center}
\includegraphics[width=10cm]{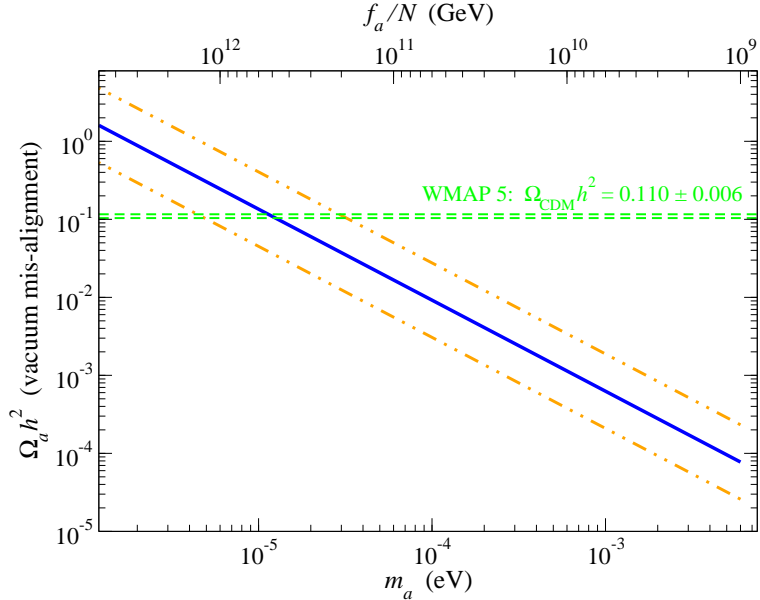}
\end{center}
\vspace{-3mm}
\caption{\small\it The expected axion relic density due to vacuum
mis-alignment versus $m_a$ (lower scale) and $f_a/N$ (upper scale). The
dashed-double-dotted lines show the typical error band on this estimate,
while the horizontal band shows the WMAP5 CDM measured abundance. 
Figure is from Ref. \cite{axino}. }
\label{fig:axion} 
\end{figure}

The expected axion relic density from vacuum mis-alignment, along with 
typical error bands, 
is shown in Fig. \ref{fig:axion}. It is worth remembering
that there is a small chance that 
$\langle a(t)\rangle \ll f_a$, in which case much lower values
of relic density could be obtained. Additional entropy production
at $t>t_{QCD}$ can also lower the axion relic abundance.  Taking the
value of Eq.~(\ref{eq:axrelic}) literally, and comparing to the WMAP5
measured abundance of CDM in the universe, one gets 
a lower bound $m_a\agt 10^{-5}$
eV on the axion mass, and a corresponding 
upper bound
$f_a/N\alt 5\times 10^{11}$ GeV, on the axion decay constant. 
%

Relic axion search experiments such as ADMX are ongoing. In these
experiments, one mounts a super-cooled microwave cavity, and searches
for relic axion scattering off microwave photons to yield photons with
energy equal to the axion mass. Only recently have experiments begun to
probe the theoretically favored regions of $f_a/N$. A thorough search is
expected to continue over the next 5-10 years\cite{admx}

\subsection{Mixed axion/axino warm and cold dark matter}

If we adopt the MSSM as the effective theory below $M_{\rm GUT}$, and
also implement a solution to the strong $CP$ problem via the PQ
mechanism, we must introduce not only an axion but also a spin-${1\over
2}$ {\it axino} $\ta$ into the theory.  The axino mass is very
model-dependent, and can be anywhere in the range of
keV-GeV\cite{axmass}. Its coupling is suppressed by the Peccei-Quinn
breaking scale $f_a$, which is constrained to be of order $10^9-10^{12}$
GeV: thus, the axino interacts more weakly than a WIMP, but not as
weakly as a gravitino.  The axion/axino mixture can be a compelling
two-component choice for DM in the universe.

Like the gravitino, the axino will likely not be in thermal equilibrium
in the post-inflation era. But it can still be produced thermally via particle
scattering. Its abundance via thermal production is given
by \cite{roszk,relic_axino}, 
\be 
\Omega_{\ta}^{TP}h^2  \simeq  5.5
g_s^6\log\left(\frac{1.108}{g_s}\right) \left(\frac{10^{11}\ {\rm
GeV}}{f_a/N}\right)^2\left(\frac{m_{\ta}}{100\
{\rm MeV}}\right)\left(\frac{T_R}{10^4\ {\rm GeV}}\right)\;,
\ee 
%
where $g_s$ is the strong coupling at the reheating scale.
The axino can also be produced non-thermally by NLSP
decays, and so will inherit the thermally produced NLSP number density. 
The total axino abundance is thus given by, 
\be
\Omega_{\ta}h^2=
\Omega_{\ta}^{TP}h^2+\frac{m_{\ta}}{m_{NLSP}}\Omega_{\rm NLSP}h^2 .
\ee
Thermally produced axinos will be CDM for $m_{\ta}\agt
0.1$ MeV\cite{roszk}, while the axinos produced in NLSP decay will constitute
hot/warm DM for $m_{\ta}\alt 1$ GeV\cite{jlm}. 
Since the PQ scale is considerably lower than the Planck
scale, the lifetime for decays such as $\tz_1\to \gamma \ta$ are of
order $\sim 0.01-1$ sec -- just before BBN. Thus, the axino DM scenario is
 less constrained by BBN than gravitino DM\cite{roszk}. 

Note also that if axinos are the CDM of the universe, then models with
very large thermal neutralino abundance $\Omega_{\tz_1}h^2\sim 100-1000$ can be
readily accommodated, since there is a huge reduction in relic density
upon $\tz_1$ decay to the axino. This possibility occurs in models with
multi-TeV scalars (and hence a multi-TeV gravitino) and a bino-like
$\tz_1$.  In this case with very large $m_{\tG}$ there is no gravitino
problem as long as the re-heat temperature $T_R\sim 10^6-10^8$ GeV.
This range of $T_R$ is also what is needed to obtain successful {\it
non-thermal} leptogenesis (involving heavy neutrino $N$ production via
inflaton decay) \cite{ntlepto} along with the correct abundance of axino
dark matter \cite{axino}.

\subsubsection{Yukawa-unified SUSY with mixed axion/axino dark matter:}
Supersymmetric models wherein the $t-b-\tau$ Yukawa couplings are
unified at $Q=M_{GUT}$ are highly motivated by simple SUSY GUT models
based on the gauge group $SO(10)$. In addition, these models provide a
natural explanation for $R$-parity conservation of renormalizable
interactions, and easily accommodate see-saw neutrino masses.
Explicit RGE calculations within the MSSM find that Yukawa unification
can only occur for very precise soft SUSY breaking boundary conditions
\cite{raby,yukus,bkss}: matter scalars have mass $m_{16}\sim 10$ TeV while
the GUT scale $A_0$ terms and Higgs scalars are related as
$4m_{16}^2=2m_{10}^2=A_0^2$. With $m_{1/2}$ as small as possible, and
$\tan\beta\sim 50$, such models predict first and second generation
matter scalars at around the 10 TeV scale, third generation scalars,
$\mu$ and $m_A$ around a few TeV, gluinos around $350-500$ GeV, and a
bino-like neutralino around $50-90$ GeV \cite{yukus,bkss}. However, these
models then predict $\Omega_{\tz_1}h^2\sim 10^2-10^4$, {\it i.e.}  3-5
orders of magnitude above the measured value.

This seemingly enormous DM relic density can be reconciled with
Yukawa-unified SUSY by extending them to include an axion/axino
supermultiplet\cite{bkss} required for
the PQ solution to the strong $CP$ problem. 
In this case, if $m_{\ta}\sim 1-100 $ MeV, then
the factor $m_{\ta}/m_{\tz_1}$ suppresses the relic density by the
required factor of $10^3-10^5$. The axinos produced via neutralino decay
would constitute warm DM, but the thermally produced $\ta$s and the $a$s
would constitute cold DM. It is straightforward to find Yukawa-unified
models in Ref. \cite{axino} where the bulk ($\sim 90\%$ )of DM is cold
axions and axinos, with a smaller contribution of warm non-thermal axinos. 
The large value of $m_{16}$, related to
$m_{\tG}$ under supergravity, allows for a solution to the gravitino
problem, and allows for a re-heat temperature in the range $T_R\sim
10^6-10^8$ GeV: enough for at least a non-thermal leptogenesis solution
to the baryogenesis problem. In this scenario, WIMP DD and ID detection
experiments will likely have null results. However, a
detectable  axion signal may
be possible.  In addition, with $m_{\tg}\sim 350-500$ GeV, 
SUSY signals containing multiple isolated leptons, jets
and missing $E_T$ should soon be visible at LHC\cite{lhcso10}.
In fact, early detection of these light gluinos should be possible via
isolated multi-lepton plus multi-jet searches, even before 
$\eslt$ becomes a reliable cut variable\cite{multilep}.

\section{Summary and Outlook}

Science has entered into an era of unprecedented interaction between particle
physics, astrophysics and cosmology.  It is now certain that the bulk of
the matter in the universe is cold and non-luminous: it is 
not made of any of the known particles, 
but instead must be made of one or more {\it new matter states} 
associated with {\it physics beyond the SM}. 
Many new physics theories which
address the mechanism behind electroweak symmetry breaking and the
stabilization of the electroweak scale
naturally contain a stable WIMP particle which may serve as a 
natural candidate for the observed dark matter. 
In this review, we have focused our attention on
what we believe is the most compelling of these suggestions: {\it weak scale
supersymmetry}, which provides a phenomenologically viable, perturbatively
calculable framework with the  strong and electroweak gauge
interactions unified in a straightforward way.

SUSY theories with a conserved $R$-parity always contain a stable
particle that in many models has the right properties to be cold
dark matter.  Within the much-studied mSUGRA model discussed in
Sec.\ref{sec:sugra}, the neutralino relic density is typically too large
over most of the parameter space. There are, however, special regions,
mostly at the edge of the $m_0-m_{1/2}$ plane, where the neutralino
annihilation rate in the early universe is enhanced, bringing its
predicted {\it thermal} relic density in accord with the measured
CDM density. However, in various extensions of mSUGRA
where the underlying scalar/gaugino mass universality is relaxed by the
introduction of just one additional parameter, this is no longer the
case and, in fact, the entire $m_0-m_{1/2}$ plane is compatible with
the relic density measurement. This calls into question implications of
the relic-density-measurement for collider and other SUSY searches from
an analysis of just the mSUGRA framework. However, as discussed at the end
of Sec.~\ref{sec:sugra}, some more robust conclusions applicable to a
wide class of gravity-mediated SUSY breaking models may be possible.

We have also examined prospects for direct and indirect detection of
DM. An exciting aspect is that a wide variety of models with MHDM have a
DD cross section $\sigma_{\rm SI}(\tz_1p) \agt 10^{-8}$~pb, just an
order of magnitude away from current limits, and accessible to the next
generation of detectors, {\it e.g.} Xenon-100/LUX: see Fig.~\ref{fig:dd}. These
models may also lead to observable signals in the IceCube experiment,
and perhaps, also via other ID experiments.
Depending on the underlying theoretical reason for the small value of
$\mu^2$ needed for MHDM, there will be different implications for LHC
and ILC experiments \cite{wtnreview}. The  message here is that collider
experiments, in tandem with direct and indirect searches, will serve to
reveal the underlying physics. 
A truly unprecedented feature of this 
program is that if the SUSY WIMP composes the bulk of DM,
measurements of the properties of associated  sparticles produced at the
LHC (possibly complemented by measurements at an electron-positron
linear collider) may allow us to independently infer just how much DM
there is in the universe, and quantitatively predict what other searches
for DM should find. If these predictions turn out to be in agreement with
observation, we would have direct evidence that DM mostly consists of just a
single component.

While thermal WIMPs provide the simplest and most economic model of DM,
it is possible that the DM consists of a particle with interactions that
are so weak that it has never been in thermal equilibrium since it was
produced during the re-heating phase in the post-inflation era. In this
case, the evaluation of the relic density is more complicated and
depends on (unknown) details of the thermal history of the Universe. 
The gravitino, considered in Sec.~\ref{sec:gravitino}, 
is a viable DM candidate that has only
gravitational interactions. Indeed, the advocates of gravitino DM have argued
that the scenario solves some of the potential problems associated with
Big Bang nucleosynthesis. In these models, direct and indirect detection
experiments should have null results. A long-lived neutralino NLSP will
escape a collider detector, resulting in $\eslt$ events, while a charged
(or colored) NLSP will manifest itself as a charged penetrating track
of a slow-moving particle, together with a smoking-gun late-decay signature in
dedicated searches\cite{feng2}. Mixed axion/axino DM, considered in
Sec.~\ref{sec:axion}, is another possibility for DM with superweak
interactions. An axion signal would be the only DM signature in such a
scenario, though as for the gravitino LSP scenario, a charged NLSP would
readily reveal itself at the LHC. It has recently been argued that
Yukawa-unified models with very light gauginos but very heavy scalars,
well beyond the reach of the LHC, augmented by an axion/axino
supermultiplet are compatible with the measured value of DM while
solving the gravitino problem common to all SUGRA models.

With the LHC scheduled to begin accumulating a significant amount of
data starting in 2010, and with many direct and indirect
DM detection searches already underway, we are entering what we hope will be a
data-driven era of particle physics and cosmology.
The new noble liquid detectors for direct DM detection are already
competitive with solid-state detectors. Both approaches are beginning to
probe highly motivated regions of model parameter space, and very soon
should be able to discover (or exclude) DM from a wide class of models
with MHDM.  As discussed in Sec.~\ref{sec:indirect}, there already exist
several hints of signals from ID detection experiments.  Although it may
be likely that these hints turn out to be spurious, the important thing
is that these new probes are beginning to explore new regimes and the
new, more powerful set-ups such as the FGST should begin to provide
precision data very soon.  If a definitive WIMP signal emerges from
collider and/or direct detection experiments, then it is possible that
ID searches will not only corroborate this, but will also serve
to map out the galactic DM halo profile. Experiments over the next 10-15
years will likely reveal the identity of dark matter.  There is no doubt
that the unprecedented synthesis of the physics of the largest and
smallest scales observable in nature will make the next two decades very
exciting!

\section* {Acknowlegements} This research was supported in part
  by the United States Department of Energy. EKP was supported by the Marie
  Curie Research and Training Network {\it UniverseNet} under contract
  no. MRTN-CT-2006-035863.

\label{sec:conclude}


\section*{References}

\begin{thebibliography}{10}
%
\bibitem{zwicky} F.~Zwicky, {\em Helvetica Physica Acta} {\bf 6} (1933) 110;
  see also \apj{86}{1937}{217}.
%
\bibitem{rubin} V.~Rubin and W.~K.~Ford, \apj{159}{1970}{379};
  V.~Rubin, N.~Thonnard  and W.~K.~Ford, \apj{238}{1980}{471}.
%
\bibitem{wmap} D.~N.~Spergel {\it et al.} (WMAP Collaboration), 
{\em Astrophys.~J.~Supp.}, {\bf 170} (2007) 377
%
\bibitem{lense} J. A. Tyson, G. P. Kochanski and I. P. Dell'Antonio,
\apj{498}{1998}{L107}; H. Dahle, \astroph{0701598};
D. Clowe {\it et al.}, \astroph{0608407}.
%
\bibitem{nbody} M. Tegmark {\it et al.} (SDSS collaboration),
\apj{606}{2004}{702}.
%
\bibitem{comp} For a review, see O. Lahav and A. R. Liddle,
in {\it Review of Particle Physics}, \plb{667}{2008}{1}.
%
\bibitem{accel} A.~Riess {\it et al.} Astro. J. {\bf 116} (1998) 1009;
  S.~Perlmutter {\it et al.} \apj{517}{1999}{565}.
%
\bibitem{weinberg_cc} S. Weinberg, \prl{59}{1987}{2607}. 
%
\bibitem{cdm_reviews} 
For reviews, see {\it e.g.}
G.~Jungman, M.~Kamionkowski and K.~Griest,\prep{267}{1996}{195};
A.~Lahanas, N.~Mavromatos and D.~Nanopoulos, \ijmpd{12}{2003}{1529};
M.~Drees, \hepph{0410113};
K.~Olive, ``Tasi Lectures on Astroparticle Physics'', \astroph{0503065};
G. Bertone, D. Hooper and J. Silk, \prep{405}{2005}{279}.
For a recent review of axion/axino dark matter, see
F. Steffen, arXiv:0811.3347 (2008).
%
\bibitem{fk} See {\it e.g.} J.~Feng and J.~Kumar, \prl{101}{2008}{231301}

%
\bibitem{pqww} R. Peccei and H. Quinn, \prl{38}{1977}{1440} and
\prd{16}{1977}{1791}; S. Weinberg, \prl{40}{1978}{223};
F. Wilczek, \prl{40}{1978}{279}.
%
%
\bibitem{wss} For text book reviews, see, H.~Baer and X.~Tata, {\it Weak
Scale Supersymmetry: From Superfields to Scattering Events}, (Cambridge
University Press, 2006); M. Drees, R. Godbole and P. Roy,{\it Theory and
Phenomenology of Sparticles}, (World Scientific, 2004); P. Bin´etruy,
{\it Supersymmetry} (Oxford University Press, 2006).
%
\bibitem{feng} J. Feng, A. Rajaraman and F. Takayama, \prl{91}{2003}{011302}
and \prd{68}{2003}{063504}.
%
\bibitem{absik} L. F. Abbott and P. Sikivie, \plb{120}{1983}{133};
J. Preskill, M. Wise and F. Wilczek, \plb{120}{1983}{127};
M. Dine and W. Fischler, \plb{120}{1983}{137};
M. Turner, \prd{33}{1986}{889}
%
\bibitem{wilc} K. Rajagopal, M. Turner and F. Wilczek, 
\npb{358}{1991}{447}.
%
\bibitem{roszk} L. Covi, J. E. Kim and L. Roszkowski, \prl{82}{1999}{4180}; 
L. Covi, H. B. Kim, J. E. Kim and L. Roszkowski, \jhep{0105}{2001}{033};
L. Covi, L. Roszkowski and M. Small, \jhep{0207}{2002}{023}.
%
\bibitem{moroi_rhsn} L. Hall, T. Moroi and H. Murayama, 
\plb{424}{1998}{305}; T. Asaka, K. Ishiwata and T. Moroi, 
\prd{73}{2006}{051301} and \prd{75}{2007}{065001}.
%
%
\bibitem{isajet} ISAJET, by H.~Baer, F.~Paige, S.~Protopopescu and
X.~Tata, \hepph{0312045}; see also
H.~Baer, J.~Ferrandis, S.~Kraml and W.~Porod, \prd{73}{2006}{015010}.
%
\bibitem{isared} IsaRED, by H.~Baer, C.~Balazs and A.~Belyaev,
\jhep{0203}{2002}{042}.
%
\bibitem{msugra} 
A.~Chamseddine, R.~Arnowitt and P.~Nath, \prl{49}{1982}{970};
R.~Barbieri, S.~Ferrara and C.~Savoy, \plb {119}{1982}{343};
N.~Ohta, Prog. Theor. Phys. {\bf 70} (1983) 542; L. Hall,
J. Lykken and S. Weinberg, \prd {27}{1983}{2359}.
%
\bibitem{gg} G.~Gelmini and P.~Gondolo, \prd {74}{2006}{023510}.
%
\bibitem{bulk}H.~Goldberg, \prl {50}{1983}{1419};
J.~Ellis {\it et al.} \npb {238}{1984}{453}; P.~Nath and R.~Arnowitt,
\prl {70}{1993}{3696}; H.~Baer and M.~Brhlik, \prd{53}{1996}{597};
V.~Barger and C.~Kao, \prd{57}{1998}{3131}.
%
\bibitem{stau} J.~Ellis, T.~Falk and K.~Olive, \plb{444}{1998}{367}; 
J.~Ellis, T.~Falk, K.~Olive and M.~Srednicki, \app{13}{2000}{181};
M.E.~G\'{o}mez, G.~Lazarides and C.~Pallis, \prd{61}{2000}{123512}
and \plb{487}{2000}{313};
A.~Lahanas, D.~V.~Nanopoulos and V.~Spanos, \prd{62}{2000}{023515};
R.~Arnowitt, B.~Dutta and Y.~Santoso, \npb{606}{2001}{59}; 
see also Ref.~\cite{isared}.
%
\bibitem{hb_fp} K.~L.~Chan, U.~Chattopadhyay and P.~Nath, \prd{58}{1998}{096004};
J.~Feng, K.~Matchev and T.~Moroi, \prl{84}{2000}{2322} and 
\prd{61}{2000}{075005}; see also 
H.~Baer, C.~H.~Chen, F.~Paige and X.~Tata, \prd{52}{1995}{2746} and 
\prd{53}{1996}{6241}; 
H.~Baer, C.~H.~Chen, M.~Drees, F.~Paige and X.~Tata, \prd{59}{1999}{055014}; 
for a model-independent approach, see
H.~Baer, T.~Krupovnickas, S.~Profumo and P.~Ullio, \jhep{0510}{2005}{020}.
%
\bibitem{Afunnel}M.~Drees and M.~Nojiri, \prd{47}{1993}{376}; 
H.~Baer and M.~Brhlik, \prd{57}{1998}{567};
H.~Baer, M.~Brhlik, M.~Diaz, J.~Ferrandis, P.~Mercadante,
P.~Quintana and X.~Tata, \prd{63}{2001}{015007};
J.~Ellis, T.~Falk, G.~Ganis, K.~Olive and M.~Srednicki, \plb{510}{2001}{236}; 
L.~Roszkowski, R.~Ruiz de Austri and T.~Nihei, \jhep{0108}{2001}{024}; 
A.~Djouadi, M.~Drees and J.~L.~Kneur, \jhep{0108}{2001}{055}; 
A.~Lahanas and V.~Spanos, \epjc{23}{2002}{185}.
%
\bibitem{hfunnel} R.~Arnowitt and P.~Nath, \prl{70}{1993}{3696};
H.~Baer and M.~Brhlik, Ref.~\cite{bulk};  
A.~Djouadi, M.~Drees and J.~Kneur, \plb{624}{2005}{60}.
%
\bibitem{stop_co}C.~B\"ohm, A.~Djouadi and M.~Drees,
  \prd{62}{2000}{035012}; 
J.~R.~Ellis, K.~A.~Olive and Y.~Santoso, \app{18}{2003}{395};
J.~Edsj\"o, {\it et al.}, JCAP {\bf 0304} (2003) 001.
%
\bibitem{isares} H.~Baer, C.~Balazs,
A.~Belyaev and J.~O'Farrill, JCAP {\bf 0309}, (2003) 007.
%
\bibitem{dsusy} P.~Gondolo, J.~Edsjo, P.~Ullio, L.~Bergstrom, 
M.~Schelke and E.~A.~Baltz, JCAP {\bf 0407} (2004) 008.
%
\bibitem{bo}H.~Baer and J.~O'Farrill, JCAP {\bf 0404} (2004) 005;
H.~Baer, A.~Belyaev, T.~Krupovnickas and J.~O'Farrill, 
JCAP {\bf 0408} (2004) 005.
%
\bibitem{pulsars} D.~Hooper, P.~Blasi and P.~Serpico, JCAP {\bf 0901}
  (2009) 025; S.~Profumo, arXiv:0812.4457 [astro-ph]; H.~Yuksel,
  M.~Kistler and T.~Stanev, arXiv:0810.2784 [astro-ph].
%
\bibitem{aandm} R.~Arnowitt {\it et al.} \plb {649}{2007}{73} and \prl
  {100}{2008}{231802}. 
%
\bibitem{lhcreach} H.~Baer, C.~Bal\'azs, A.~Belyaev, T.~Krupovnickas and X.~Tata,
\jhep{0306}{2003}{054}; see also,  
S.~Abdullin and F.~Charles, \npb{547}{1999}{60};
S.~Abdullin {\it et al.} (CMS Collaboration), \jphg{28}{2002}{469} [\hepph{9806366}];
B.~Allanach, J.~Hetherington, A.~Parker and B.~Webber, 
\jhep{08}{2000}{017}.
%
\bibitem{mizukoshi} J.~Mizukoshi, P.~Mercadante and X.~Tata,
  \prd{72}{2005}{035009}; S.~P.~Das, A.~Datta, M.~Guchait, M.~Maity and
  S.~Mukherjee, \epjc {54}{2008}{645}; R.~Kadala, J.~Mizukoshi,
  P.~Mercadante and X.~Tata, \epjc {56}{2008}{511}.
%
\bibitem{ilcreach} H. Baer, A. Belyaev, T. Krupovnickas and   X. Tata, 
\jhep{0402}{2004}{007};
H.~Baer, T.~Krupovnickas and X.~Tata, \jhep{0406}{2004}{061}.
%
\bibitem{bbpw} E.~Baltz, M.~Battaglia, M.~Peskin and T.~Wizansky,
\prd{74}{2006}{103521}.  R.~Arnowitt {\it et al.},
Phys. Rev. Lett. Ref.\cite{aandm};
M.~Nojiri, G.~Polesello, D.~Tovey, \jhep{0603}{2006}{063}.
%
\bibitem{soni} S.~Soni and H.~Weldon, \plb {126}{1983}{215}
%
\bibitem{nonunigaug} C.~Hill, \plb{135}{1984}{47}; J.~Ellis {\it et al.}
  \plb{155}{1985}{381}; M.~Drees, \plb {158}{1985}{409}; See G.~Anderson
  {\it et al.} \prd {61}{2000}{095005} for the collider phenomenology of
  such models. 
%
\bibitem{wtnreview} H.~Baer, A.~Mustafayev, E.~Park and X.~Tata, \jhep
  {0805}{2008}{058}. 
%
\bibitem{wtn} H.~Baer, A.~Mustafayev, E.~Park and X.~Tata,
JCAP {\bf 0701}, 017 (2007).
%
\bibitem{pran} D. Feldman, Z. Liu and P. Nath, \plb{662}{2008}{190}.
%
\bibitem{green} R. Schnee, (CDMS Collaboration); 
A. M.~Green, JCAP {\bf 0708} (2007) 022; C-L.~Shan and
  M.~Drees, arXiv:0710.4296 [hep-ph].
%
\bibitem{bmm} V.~Barger, D.~Marfatia and A.~Mustafayev,
  \plb{665}{2008}{242}, and A.~Mustafayev (private communication).
%
\bibitem{heat} S. W. Barwick {\it et al.} (HEAT collaboration), 
\apj{482}{1997}{L191}.
%
\bibitem{willem} W.~de Boer, M.~Herold, C.~Sander, V.~Zhukov, A.~V.~Gladyshev and D.~I.~Kazakov,
  arXiv:astro-ph/0408272.
%
\bibitem{willem2} W. deBoer, C. Sander, V. Zhukov, A. Gladyshev and D. Kazakov, \plb{636}{2006}{13}.
%
\bibitem{bbs} H. Baer, A. Belyaev and H. Summy, \prd{77}{2008}{095013}.
%
\bibitem{mosk} I. Moskalenko, talk at CERN ENTApP meeting (Jan 2009).
%
\bibitem{egret2} D. Elsaesser and K. Mannheim, \prl{94}{2005}{171302}.

%
\bibitem{fink} D. P. Finkbeiner, \apj{614}{2004}{186} and \astroph{0409027}.
%
\bibitem{pamela} O.~Adriani {\it et al.} arXiv:0810.4995 (2008).
%
\bibitem{atic} J.~Chang {\it et al.} Nature {\bf 456} (2008) 362.

%
%

\bibitem{gravitinop} S. Weinberg, \prl{48}{1982}{1303};
R. H. Cyburt, J. Ellis, B. D. Fields and K. A. Olive, \prd{67}{2003}{103521};
K. Jedamzik, \prd{70}{2004}{063524};
M. Kawasaki, K. Kohri and T. Moroi, \plb{625}{2005}{7} and
\prd{71}{2005}{083502}; K. Kohri, T. Moroi and A. Yotsuyanagi, \prd{73}{2006}{123511}.
%
\bibitem{moroi2} M.~Kawasaki, K.~Kohri, T.~Moroi and A.~Yotsuyanagi,
\prd {78}{2008}{065011}.
%
\bibitem{lowscale} Low scale baryogenesis mechanisms have been discussed
  by, S.~Dimopoulos and L.~Hall, \plb{196}{1987}{135};
  J.~Cline and S.~Raby, \prd{43}{1991}{1781}; K.~Babu,
  R.~Mohapatra, and S.~Nasri, \prl{97}{2006}{131301};
  For electroweak baryogenesis, see: M.~Carena, M.~Quiros, A.~Riotto
  A.~Vilja and C.~Wagner, \npb{503}{1997}{387};
  C.~Bal\'azs {\it et al.}  \prd{71}{2005}{075002} and
  references therein; T.~Konstandin, T.~Prokopec, M.~Schmidt and
  M.~Seco, \npb{738}{2006}{1}; D.~Chung {\it et al.}
  arXiv: 0808.1144 [hep-ph] (2008).


%
\bibitem{primack} H. Pagels and J. Primack, \prl{48}{1982}{223}.
%
%
\bibitem{buchmuller} W. Buchmuller, P. Di Bari and M. Plumacher, 
{\em Annal. Phys.} {\bf 315} (2005) 305.
%
\bibitem{buchm} W. Buchmuller, L. Covi, J. Kersten, K. Schmidt-Hoberg,
JCAP {\bf 0611} (2006) 007; W. Buchmuller, L. Covi, K. Hamaguchi, A. Ibarra
and T. Yanagida, \jhep{0703}{2007}{037}.
%
\bibitem{gmodels} For some recent papers, see {\it e.g.}
J. Ellis, K. Olive, Y. Santoso and V. Spanos, \plb{588}{2004}{7} and
L. Roszkowski, R. Ruiz de Austri and K.Y. Choi, 
\jhep{0508}{2005}{080}.
%
\bibitem{thermal_G} M. Bolz, A. Brandenburg and W. Buchmuller,
\npb{606}{2001}{518}; J. Pradler and F. Steffen, \hepph{0608344}.
%
\bibitem{jlm} K. Jedamzik, M. Lemoine and G. Moultaka,
JCAP {\bf 0607} (2006) 010.
%
\bibitem{fst} J.~Feng, S.~Su and F.~Takayama, \prd{70}{2004}{075019}.
%
\bibitem{slow} M.~Drees and X.~Tata, \plb{252}{1990}{695};
 H.~Baer, K.~Cheung and J.~Gunion, \prd{59}{1999}{075002};
 M.~Fairbairn {\it et al.}  \prep{438}{2007}{1} and
 references therein. For the feasability of these searches at the LHC
 see, S.~Giagu, ATL-PHYS-PROC-2008-029 (2008); M.~Johansen, {\em Acta
 Physica Polonica}, {\bf 38} (2007) 591.
%
\bibitem{feng2} J.~Feng and B.~Smith, \prd{71}{2005}{015004} and \prd
  {71}{2005}{019904} (Erratum); K.~Hamaguchi, M.~Nojiri and A.~de~Roeck,
  \jhep {0703} {2007}{046}.
%
\bibitem{axreview} For recent reviews on axion physics, see
J. E. Kim and G. Carosi, arXiv:0807.3125 (2008);
P. Sikivie, \hepph{0509198};
M. Turner, \prep{197}{1990}{67}.
%
\bibitem{admx} L. Duffy {\it et al.}, \prl{95}{2005}{091304} and \prd{74}{2006}{012006};
for a review, see S. Asztalos, L. Rosenberg, K. van Bibber, P. Sikivie
and K. Zioutas, \arnps{56}{2006}{293}.
%
\bibitem{axmass} E. J. Chun, J. E. Kim and H. P. Nilles, 
\plb{287}{1992}{123}.
%
\bibitem{relic_axino} A. Brandenburg and F.~Steffen,
JCAP {\bf 0408} (2004) 008.
%
\bibitem{ntlepto} G. Lazarides and Q. Shafi, \plb{258}{1991}{305};
K. Kumekawa, T. Moroi and T. Yanagida, \ptp{92}{1994}{437};
T. Asaka, K. Hamaguchi, M. Kawasaki and T. Yanagida, \plb{464}{1999}{12}.
%
\bibitem{axino} H. Baer and H. Summy, \plb{666}{2008}{5};
H. Baer, S. Kraml, M. Haider, S. Sekmen and H. Summy, 
JCAP {\bf 0902} (2009) 002.
%
\bibitem{raby} T.~Blazek, M.~Carena, S.~Raby and C.~Wagner, \prd
{56}{1997}{6919}; T.~Blazek, R.~Dermisek and S.~Raby, \prd
{65}{2002}{115004}; R.~Dermisek and S.~Raby, \plb {622}{2005}{327};
R.~Dermisek, M.~Harada and S.~Raby, \prd {74}{2006}{035011};
W. Altmannshofer, D. Guadagnoli, S. Raby and D. Straub, 
\plb{668}{2008}{385}.
%
\bibitem{yukus} H.~Baer {\it et al. } \prd {61}{2000}{111701}; \prd
  {63}{2001}{015007}; D.~Auto {\it et al.} \jhep {0306}{2003}{023}.
%
\bibitem{bkss} H. Baer, S. Kraml, S. Sekmen and H. Summy, \jhep{0803}{2008}{056}.
%
\bibitem{lhcso10} H. Baer, S. Kraml, S. Sekmen and H. Summy,
\jhep{0810}{2008}{079}.
%
\bibitem{multilep} H. Baer, H. Prosper and H. Summy, \prd{77}{2008}{055017};
H. Baer, A. Lessa and H. Summy, arXiv:0809.4719 (2008).
%
\end{thebibliography}
\end{document}